\documentclass[aps,twocolumn,groupedaddress,showpacs]{revtex4}

\usepackage{amssymb,amsbsy}
\usepackage{graphicx}
\def\be{\begin{equation}}
\def\ee{\end{equation}}
\def\bea{\begin{eqnarray}}
\def\eea{\end{eqnarray}}
\def\bear{\begin{array}}
\def\ear{\end{array}}
\def\bfig{\begin{figure}}
\def\efig{\end{figure}}
\def\bcen{\begin{center}}
\def\ecen{\end{center}}
\def\raw{\rightarrow}

\def\vp{\mathbf{p}}

\def\la{\label}
\def\chic{\scriptscriptstyle}

\def\D{\displaystyle}
 
\def\bphi{\boldsymbol\phi}

\def\bnu{\boldsymbol\nu}

\begin{document}


\title{$\bphi$ meson propagation in a hot hadronic gas}

\author{L. Alvarez-Ruso}
\author{V. Koch}
\affiliation{Nuclear Science Division, Lawrence Berkeley National Laboratory
\\ 1 Cyclotron Rd., Berkeley, CA 94720, USA}


\begin{abstract}
The Hidden Local Symmetry Lagrangian is used to study the interactions of
$\phi$ mesons with other pseudoscalar and vector mesons in a hadronic gas at
finite temperature. We have found a significantly small $\phi$ mean free
path (less than 2.4~fm at $T > 170$~MeV) due to large collision rates with
$\rho$ mesons, kaons and predominantly $K^*$ in spite of their heavy mass.
This implies that $\phi$ mesons
produced after hadronization in relativistic heavy ion collisions will not
leave the hadronic system without scattering. The effect of these
interactions on the time evolution of the $\phi$ density in the expanding
hadronic fireball is investigated.           
\end{abstract}

\maketitle

\section{Introduction}

Enhanced strangeness production in relativistic
nucleus-nucleus collisions compared to nucleon-nucleon collisions has been predicted to be a
consequence of the formation of quark-gluon plasma
(QGP)~\cite{Rafelski:1982pu}. This effect leads to an enhancement in the
number of strange and multistrange particles produced after hadronization. In
particular, free $s \bar{s}$ pairs would coalesce to form $\phi$ mesons,
while their production in $pp$ and $\pi p$ collisions should be OZI
suppressed~\cite{Shor:1985ui}. On the other side, the $\phi$ mass is expected
to decrease due to both many body effects in the hadronic
medium~\cite{Asakawa:1994tp, Song:1996gw} and chiral symmetry
restoration, which might generate  a double peak structure~\cite{Asakawa:1994nn}.
The decay width is also modified (broadened) through the scattering with
other particles~\cite{Ko:1994id, Haglin:1995xu, Smith:1998xu}. 

The $\phi$ is
a nice probe since it is not masked behind other resonances in the mass
spectra. It decays into kaon pairs ($K^+ K^-$), and more rarely
into dileptons ($e^+ e^-$, $\mu^+ \mu^-$); both channels have been detected
at CERN-SPS~\cite{Afanasiev:2000yo,Abreu:1996yg,Abreu:2001qp}.
Dileptons have negligible final state
interactions with the hadronic environment, so they sense the entire
evolution of the system. On the contrary, detectable kaons from $\phi$ decay
probably emerge only at freezeout. 

It is widely accepted that the $\phi$
mean free path (MFP) in the hot hadronic fireball is large 
due to the small cross
section for scattering with nonstrange hadrons. This implies that $\phi$
spectra would retain the information about the stage of the collision at
which the plasma hadronizes~\cite{Shor:1985ui}. Available
calculations~\cite{Ko:1994id, Haglin:1995xu, Smith:1998xu} seem to support
this idea. For example, in Ref.~\cite{Haglin:1995xu} the $\phi$ MFP
was calculated taking into account its scattering with different
mesons. Phenomenological Lagrangians with couplings extracted from the
experimental partial decay rates were used. The obtained MFP at $T=200$~MeV
is rather big ($\lambda = 4.4$~fm) compared
to the standard size of the hadronic system. Adding
the reactions with nucleons and nucleonic resonances did not change
qualitatively the situation~\cite{Smith:1998xu}.     

However, $\phi$ production in $Pb-Pb$ collisions at SPS shows some intriguing
features that are difficult to match with the picture of a $\phi$  weakly
interacting with the hadronic medium. Both absolute yields and inverse slope
parameters in the transverse mass ($m_t$)  distributions exhibit different values
when measured via $\mu^+ \mu^-$ or $K^+ K^-$
decays~\cite{Rohrich:2001qi}. The inverse slope, as obtained from the
hadronic measurement, suggests that the $\phi$'s flow together with pions,
kaons and protons, while the dilepton measurement is consistent with the
assumption of an early freezeout. It has also been observed that the rapidity
distribution (extracted from kaon pairs) in $Pb-Pb$ is about 50~\%
broader than in $pp$~\cite{Afanasiev:2000yo}. The origin could be attributed to
longitudinal flow~\cite{Safarik:2001qy}. The modification of visible $\phi$
spectra due to kaon rescattering inside the fireball is an important
ingredient but
does not fully explain the discrepancies. Indeed, Johnson et
al.~\cite{Johnson:1999fv} show that kaon rescattering can account for 
the observed rapidity distributions, but not for the differences in the
$m_t$ spectra, neither the inverse slopes, nor the relative yields. 

Here we calculate $\phi$ collision rates and mean free path in a hot
hadronic gas of pseudoscalar ($\pi$, $K$) and vector mesons ($\rho$,
$\omega$, $K^*$, $\phi$). The reaction cross sections are obtained within the Hidden Local
Symmetry Lagrangian (HSL)~\cite{Bando:1988br}, which includes both Goldstone bosons and vector mesons
in a manner consistent with the  symmetries of QCD. The use of such a
realistic model allows us to take into account many mechanisms that are
not present in calculations that rely only in couplings extracted from observed
decays but are allowed by the symmetries. This is, for instance, the case of
vertices like $\phi K^* K$, $\rho K^* K^*$ and many others. As a consequence,
we shall see that at temperatures between 150 and 200~MeV the
$\phi$ mean free path in hadronic matter is considerably smaller than what
has been estimated so far. Finally, we study the implications of these 
findings for the $\phi$ yields.

\section{The Hidden Local Symmetry Lagrangian}
\label{HLS}

The HLS model provides a natural framework for describing the interactions
between  pseudoscalar and vector mesons. It is based on the fact that a
[U(3)$_{\mathrm L} \times$ U(3)$_{\mathrm R}$]/U(3)$_{\mathrm V}$ non-linear sigma model is gauge
equivalent to another one with [U(3)$_{\mathrm L} \times$ U(3)$_{\mathrm
R}$]$_\mathrm{global} \times$[U(3)$_{\mathrm V}$]$_\mathrm{local}$ symmetry.
The most general Lagrangian possessing this symmetry and made with the
smallest number of derivatives is~\cite{Bando:1985rf}
\be
\la{Lin}
{\mathcal L} = {\mathcal L}_A + a {\mathcal L}_V \,,
\ee
\be
\la{LA}
{\mathcal L}_A = - \frac{1}{4} f_\pi^2\,\left\langle \left(D_\mu \xi_L\cdot\xi_L^\dagger - 
D_\mu \xi_R\cdot\xi_R^\dagger\right)^2 \right\rangle\,,
\ee
\be
\la{LV}
{\mathcal L}_V = - \frac{1}{4} f_\pi^2\,\left\langle\left(D_\mu
\xi_L\cdot\xi_L^\dagger + D_\mu \xi_R\cdot\xi_R^\dagger\right)^2 \right\rangle\,,
\ee
where $f_\pi$ stands for the pion decay constant and $a$ is an arbitrary
parameter; it is usually set to $a=2$, which allows to recover the standard
vector meson dominance expression (VMD)~\cite{Bando:1985ej,Bando:1985rf}, although a
slightly bigger value has been extracted from
experiment~\cite{Benayoun:1998ex}. The symbol $\langle\rangle$ denotes
flavor trace. The covariant derivatives of the fields
$\xi_{L,R}$ are   
\be
\la{coder}
D_\mu \xi_{L(R)} = \left( \partial_\mu - i g V_\mu \right) \xi_{L(R)} \,,
\ee
with $V_\mu$ being the dynamical gauge bosons of the hidden local symmetry,
further identified with the nonet of vector mesons 
\be
\la{vm}
\bear{l}
V_\mu \equiv V_\mu^a T^a \\[0.4cm]
\D = \frac{1}{\sqrt{2}} \left(
\bear{ccc}
\D
\frac{1}{\sqrt{2}} \rho_\mu^0 + \frac{1}{\sqrt{2}} \omega_\mu &
\rho_\mu^+ & K_\mu^{*+}\\[0.4cm]
\rho_\mu^- &\D -\frac{1}{\sqrt{2}} \rho_\mu^0 + \frac{1}{\sqrt{2}}\omega_\mu &
K_\mu^{*0}\\[0.4cm]
K_\mu^{*-} & \bar{K}_\mu^{*0} & \phi_\mu \\ 
\ear
\right)\,.
\ear
\ee

Working in the unitary gauge i.e. choosing $\xi_{L,R}$ such that 
\be
\la{ug}
\xi^\dagger_L = \xi_R = \xi =\exp{i \Phi/f_\pi}
\ee
$\mathcal{L}_\mathrm{A}$ becomes identical to the lowest order chiral
Lagrangian 
\be
\la{chpt}
\mathcal{L}_\mathrm{A} = \frac{f^2_\pi}{4}  \langle \partial_\mu U
\partial_\mu U^\dagger \rangle \,,
\ee
with $U=\xi^2$; $\Phi$ is the octet of pseudoscalar Goldstone bosons
\be
\la{psm}
\Phi = \frac{1}{\sqrt{2}} \left(
\bear{ccc}
\D \frac{1}{\sqrt{2}} \pi^0 + \frac{1}{\sqrt{6}} \eta & \pi^+ & K^+ \\[0.4cm]
\pi^- &\D -\frac{1}{\sqrt{2}} \pi^0 + \frac{1}{\sqrt{6}} \eta & K^0 \\[0.4cm]
K^- & \bar{K}^0 &\D - \frac{2}{\sqrt{6}} \eta \\
\ear
\right) \,.
\ee
The physics contained in $\mathcal{L}_V$ becomes clear when $\xi$  is
expanded up to the term linear in $\Phi$
\be
\la{exp}
\xi \approx 1 + i \frac{\Phi}{f_\pi} \,.
\ee
Then, 
\be
\la{LV1} 
{\mathcal L}_V =\frac{1}{2} a g^2 f_\pi^2 V_\mu^a V^{\mu a} 
- i \frac{a}{2} g \langle\left\{\left[\Phi,\partial_\mu \Phi \right], V^\mu
\right\} \rangle \,,  
\ee
where $[,]$ and $\{,\}$ stand for commutator and anticommutator respectively.  
Vector mesons have acquired mass $m_V^2=a g^2 f_\pi^2$  by spontaneous breakdown 
of the hidden local symmetry. One now assumes that a kinetic term  for
them is generated by the underlying QCD dynamics or quantum effects at the
composite level~\cite{Bando:1985ej, Bando:1985rf}. Thus, we have                       
\be
\la{Lkin}
{\mathcal L}_{kin} = - \frac{1}{2} \langle F_{\mu \nu}  F^{\mu \nu} \rangle \,, 
\ee
$F_{\mu \nu}$ being the nonabelian field strength tensor 
\be
\la{fmunu}
F_{\mu \nu} = \partial_\mu V_\nu - \partial_\nu V_\mu - i g
\left[V_\mu,V_\nu\right] \,.
\ee
Substituting Eq.~(\ref{fmunu}) into (\ref{Lkin}) we obtain, apart from the
standard kinetic terms, the interactions of 3 and 4 vector mesons 
\be
\la{VVV}
{\mathcal L}_{VVV} = i g  \langle (\partial_\mu V_\nu -
\partial_\nu V_\mu) \left[V^\mu,V^\nu\right] \rangle 
\ee
\be
\la{VVVV} 
{\mathcal L}_{VVVV}= - \frac{g^2}{2} \langle \left[V_\mu,V_\nu\right]
\left[V^\mu,V^\nu\right] \rangle
\ee
As we shall see, these terms generate large contributions to the total cross
section of the $\phi$ meson with other hadrons, specially vector mesons.  

In order to account for deviation from the flavor symmetry one should add
[U(3)$_{\mathrm L} \times$ U(3)$_{\mathrm R}$]$_\mathrm{global}$ breaking
terms which do not affect the hidden symmetry. There is no unique way to do
this. Different implementations are studied in Ref.~\cite{Benayoun:1998hz}.
Here we adopt the scheme proposed in Ref.~\cite{Bramon:1995cb}, in which
${\mathcal L}_A$ and ${\mathcal L}_V$ are modified as follows 
\bea
\la{LAmod}
{\mathcal L}_A + \Delta {\mathcal L}_A = 
- \frac{1}{4} f_\pi^2\, \Big\langle \left( D_\mu \xi_L\cdot\xi_L^\dagger - 
D_\mu \xi_R\cdot\xi_R^\dagger \right)^2 &\times& \nonumber \\[0.2cm]
\left( 1 + \xi_L \epsilon_A 
\xi_R^\dagger + \xi_R \epsilon_A \xi_L^\dagger \right) \Big\rangle\,,&&
\eea
\bea
\la{LVmod}
{\mathcal L}_V + \Delta {\mathcal L}_V  = - \frac{1}{4} f_\pi^2\,\left\langle \left(D_\mu
\xi_L\cdot\xi_L^\dagger + D_\mu \xi_R\cdot\xi_R^\dagger\right)^2 
\right. &\times& \nonumber \\[0.2cm]
\left. \left(1 + \xi_L \epsilon_V
\xi_R^\dagger + \xi_R \epsilon_V \xi_L^\dagger \right) \right\rangle \,,&&
\eea
with $\epsilon_{A(V)} =\mathrm{diag} (0,0,c_{\chic{A(V)}})$; $c_{\chic{A(V)}}$ are real
parameters to be determined. After fixing the gauge (Eq.~\ref{ug}) and 
expanding $\xi$ (Eq.~\ref{exp}), one observes that the kinetic terms in
${\mathcal L}_A$ have to be renormalized. This is achieved by rescaling~\cite{Bramon:1995cb} 
\be
\la{resc}
\sqrt{1+c_{\chic{A}}}\,K  \raw K\,, \quad \sqrt{1+\frac{2}{3}c_{\chic{A}}}\,\eta \raw \eta \,. 
\ee 
Such a redefinition leads to the following relations for the pseudoscalar
meson decay constants
\be
\la{fKeta}
f_K = \sqrt{1+c_{\chic{A}}}\, f_\pi \,, \quad
f_\eta=\sqrt{1+\frac{2}{3}c_{\chic{A}}}\, f_\pi \,. 
\ee
Now, different vector mesons have different masses
\be
\la{masses}
m_\rho^2 = m_\omega^2 = \frac{m_{K^*}^2}{1+c_{\chic{V}}}
=\frac{m_\phi^2}{1+2 c_{\chic{V}}}= a
f_\pi^2 g^2\,,
\ee
while the second term in Eq.~(\ref{LV1}), which describes the coupling of
vector mesons to a pair of pseudoscalar ones, looks like
\be
\la{VPP}
{\mathcal L}_{VPP} =  - i \frac{a}{2} g
\langle \left\{\left[\Phi,\partial_\mu \Phi \right], V^\mu \right\}  (1+ 2 \epsilon_V)
\rangle\,.
\ee
Eq.~(\ref{LVmod}) contains also a vector-vector-pseudoscalar-pseudoscalar
contact term
\be
\la{VVPP}
{\mathcal L}_{VVPP} = - a g^2 \left\langle V_\mu V^\mu \left( \epsilon_V
\Phi^2  + 2 \Phi \epsilon_V \Phi + \Phi^2 \epsilon_V \right) \right\rangle\,. 
\ee

Obviously, the flavor symmetry is also broken in the pseudoscalar sector.
The corresponding Lagrangian can be written as
\be
\la{SB}
{\mathcal L}_{SB} = \frac{1}{4} f_\pi^2 \left\langle 
\xi_L \chi \xi_R^\dagger + \xi_R \chi \xi_L^\dagger \right\rangle\,,
\ee
where $\chi$ can be expressed in terms of $\pi$ and $K$ masses 
\be
\chi =\mathrm{diag} (m_\pi^2, m_\pi^2, 2 m_K^2 - m_\pi^2)  
\ee
assuming the same mass for both $u$ and $d$ quarks.

It is known that the local chiral symmetry is broken at the quantum level.
The anomalous part of the Lagrangian in terms of effective degrees of
freedom is obtained assuming that the anomaly at composite level should
coincide with that at constituent level. In the framework of HLS, vector
mesons were incorporated to the anomalous Lagrangian by Fujikawa et
al.~\cite{Fujiwara:1985mp}. It contains the vector-vector-pseudoscalar
interaction. In Ref.~\cite{Bramon:1995pq}, the flavor breaking effect was
included by introducing a term $(\xi_L \epsilon_{\chic{WZ}} \xi_R^\dagger + \xi_R
\epsilon_{\chic{WZ}} \xi_L^\dagger)$, with $\epsilon_{\chic{WZ}}=\mathrm{diag}
(0,0,c_{\chic{WZ}})$, so that the total anomalous Lagrangian reads
\bea
\la{VVP}
&&{\mathcal L}_{VVP} + \Delta {\mathcal L}_{VVP}  \nonumber \\[0.2cm]
&&= 2 g_{\chic{VVP}} \epsilon^{\mu
\nu \lambda \sigma}  \left\langle \partial_\mu V_\nu \left( 1+2
\epsilon_{\chic{WZ}} \right) \partial_\lambda V_\sigma \Phi \right\rangle \,.  
\eea
The coupling constant is fixed by the anomaly
\be
\la{VVPc}
g_{\chic{VVP}}=\frac{3 g^2}{8 \pi^2 f_P}\,,\quad P=(\pi, K,\eta) \,,
\ee
and $c_{\chic{WZ}}$ is directly obtained from the ratio between the
experimental decay widths of $K^{*0} \raw K^0 \gamma$ and 
$K^{*\pm} \raw K^\pm \gamma$~\cite{Bramon:1995pq}.  

\section{Collision rates and mean free path}

\subsection{General formulae}
\la{general}

The propagation of $\phi$ mesons through the hot hadronic gas depends on how
often they interact in their way out of the fireball. For a given binary
reaction $1\,+\,2 \raw 3\,+\,4$ , the number of collisions per unit time at
a given temperature is described by the average collision rate 
\bea
\la{CRgen}
\Gamma_{coll}(T) = \frac{d_1 d_2}{n_1} \int \sum_{i=1}^{4} 
\frac{d^3 p_i}{2 E_i (2 \pi)^3} \overline{\left| \mathcal{M} \right|^2}
\times \nonumber\\[0.3cm]
(2 \pi)^4 \delta^{(4)} (p_1+p_2-p_3-p_4) f_1 f_2 (1+f_3)(1+f_4)    
\eea
where $\overline{\left| \mathcal{M} \right|^2}$ stands for the amplitude
squared, summed over final spins and averaged over initial ones; 
$d_i$ are the spin
degeneracy factors $d_i = 2S_i +1$ (we take care of the isospin degeneracy
by considering each isospin channel independently),  while $f_i$ represent 
the momentum distributions. The particle number density is given by
\be
\la{den}
n_1 = d_1 \int \frac{d^3 p_1}{(2 \pi)^3} f_1
\ee
Approximating all momentum distributions by Boltzmann 
\be
\la{bolz}
f_i\approx e^{-E_i/T}\,,\quad  1+f_i \approx 1 + e^{-E_i/T} \approx 1 \,,
\ee
and, identifying the particle labeled $1$ with the $\phi$ meson, one gets 
\bea
\la{CR}
&\Gamma_{coll}^{(2)}(T)& = \frac{d_2}{4 \pi^2 m_\phi^2 K_2(m_\phi/T)} \times
\nonumber \\[0.2cm]
&&\int_{(m_\phi+m_2)^2}^{\infty} ds \sqrt{s} p_{cm}^2 \sigma(s)
K_1(\sqrt{s}/T) 
\eea
with $s$ and $p_{cm}$ being the center-of-mass (CM) energy squared and
three-momentum respectively and $\sigma (s)$, the total cross section; 
index $(2)$ denotes the particle that collides with the $\phi$.    

The most straightforward way to estimate the relevance of rescatterings in
the medium is to compute the mean free path and compare it with the size of
the system. For the mean free path we use
\be
\la{mfp}
\lambda(T) =\frac{\bar{v}}{\Gamma_{coll}^t}\,;
\ee
$\bar{v}$ is the average velocity of the $\phi$ in the medium 
\be
\la{veloc}
\bar{v} = \frac{d_1}{n_1} \int \frac{d^3 p_1}{(2 \pi)^3} f_1
\frac{|\vp_1|}{E_1} = \frac{2 T (T+m_\phi)}{m_\phi^2 K_2(m_\phi/T)} e^{-m_\phi/T}\,;
\ee
the expression on the right is obtained assuming a Boltzmann distribution.
$\Gamma_{coll}^t = \sum_a \Gamma_{coll}^{(a)}$ arises from the contribution of
all relevant hadronic reactions. In the present calculation we have
$a=\pi,K,\rho,\omega,K^*,\phi$.  

\subsection{Binary Reactions}
\la{react}

\begin{figure}[h!]
\begin{center}
\includegraphics[height=0.25\textwidth,angle=-90]{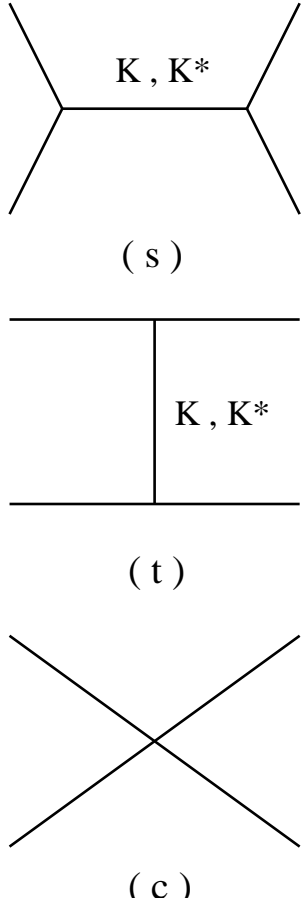}
\caption{Generic Feynman diagrams for the different reaction mechanisms; 
(s) stands for s-channel, (t) for t-channel and (c) for contact.
OZI suppressed diagrams have not been considered.}
\label{diag}
\end{center}
\end{figure}

Within the HLS model described in Section~\ref{HLS} one can compute all
possible binary reactions of $\phi$ mesons with
$\pi,K,\rho,\omega,K^*,\phi$ at tree level. The interaction vertices are given by the VPP,
VVP, VVV VVVV and VVPP Lagrangians defined in
Eqs.~(\ref{VPP}, \ref{VVP}, \ref{VVV}, \ref{VVVV}, \ref{VVPP}) respectively. 
There are
three classes of Feynman diagrams, which are depicted in Fig.~\ref{diag} and
represent s-channel (s), t-channel (t) and contact (c) reaction mechanisms.
\begin{table}[h!]
\begin{ruledtabular}
\begin{tabular}{lclcl}
No. & \hspace{0.5cm} &  Reaction & \hspace{0.5cm} &  Channels \\
\hline
1.1 & &  $\phi\, +\, \pi \raw K\, +\, K$  & & t$(K^*)$\\
1.2 & & $\phi\, +\, \pi \raw K\, +\, K^*$ & & t$(K,K^*)$, c\\
1.3 & & $\phi\, +\, \pi \raw K^*\, +\, K^*$ & & t$(K,K^*)$\\
2.1 & & $\phi\, +\, K \raw \pi\, +\, K$ & & s,t$(K^*)$\\
2.2 & & $\phi\, +\, K \raw \rho\, +\, K$ & & s,t$(K,K^*)$\\
2.3 & & $\phi\, +\, K \raw \omega\, +\, K$ & & s,t$(K,K^*)$\\
2.4 & & $\phi\, +\, K \raw \phi\, +\, K$ & & s,t$(K,K^*)$, c\\
2.5 & & $\phi\, +\, K \raw \pi\, +\, K^*$ & & s$(K,K^*)$, t$(K^*)$\\
2.6 & & $\phi\, +\, K \raw \rho\, +\, K^*$ & & s,t$(K,K^*)$\\
2.7 & & $\phi\, +\, K \raw \omega\, +\, K^*$ & & s,t$(K,K^*)$\\
2.8 & & $\phi\, +\, K \raw \phi\, +\, K^*$ & & s,t$(K,K^*)$\\
3.1 & & $\phi\, +\, \rho \raw K\, +\, K$  & & t$(K,K^*)$\\
3.2 & & $\phi\, +\, \rho \raw K\, +\, K^*$ & & t$(K,K^*)$\\
3.3 & & $\phi\, +\, \rho \raw K^*\, +\, K^*$ & & t$(K,K^*)$, c\\
4.1 & & $\phi\, +\, \omega \raw K\, +\, K$  & & t$(K,K^*)$\\
4.2 & & $\phi\, +\, \omega \raw K\, +\, K^*$ & & t$(K,K^*)$, c\\
4.3 & & $\phi\, +\, \omega \raw K^*\, +\, K^*$ & & t$(K,K^*)$, c\\
5.1 & & $\phi\, +\, K^* \raw \pi\, +\, K$ & & s$(K^*)$, t$(K,K^*)$, c\\
5.2 & & $\phi\, +\, K^* \raw \rho\, +\, K$ & & s,t$(K,K^*)$\\
5.3 & & $\phi\, +\, K^* \raw \omega\, +\, K$ & & s,t$(K,K^*)$\\
5.4 & & $\phi\, +\, K^* \raw \phi\, +\, K$ & & s,t$(K,K^*)$\\
5.5 & & $\phi\, +\, K^* \raw \pi\, +\, K^*$ & & s,t$(K,K^*)$\\
5.6 & & $\phi\, +\, K^* \raw \rho\, +\, K^*$ & & s,t$(K,K^*)$, c\\
5.7 & & $\phi\, +\, K^* \raw \omega\, +\, K^*$ & & s,t$(K,K^*)$, c\\
5.8 & & $\phi\, +\, K^* \raw \phi\, +\, K^*$ & & s,t$(K,K^*)$, c\\
6.1 & & $\phi\, +\, \phi \raw K\, +\, K$  & & t$(K,K^*)$, c\\
6.2 & & $\phi\, +\, \phi \raw K\, +\, K^*$ & & t$(K,K^*)$, c\\
6.3 & & $\phi\, +\, \phi \raw K^*\, +\, K^*$ & & t$(K,K^*)$, c\\
\end{tabular}
\end{ruledtabular}
\caption{List of binary reactions of $\phi$'s with pseudoscalar ($\pi$, $K$)
and vector ($\rho$, $\omega$, $K^*$, $\phi$) mesons. s, t and c correspond 
to the diagrams in Fig.~\ref{diag} while the particles in brackets denote
the intermediate mesons.}
\la{table}
\end{table}
The reactions considered in the present study are listed in
Table~\ref{table}. We have neglected OZI suppressed reaction mechanisms.
Therefore, only s and t diagrams with a strange particle ($K$ or $K^*$) in
the intermediate state have been considered. This is also the reason why
s-channel mechanisms are only present in the case of $\phi$ collisions with 
strange mesons. Notice that there are two kinds of contact terms: 
the VVPP vertex, which appears as a result of symmetry breaking and 
the VVVV vertex obtained from the kinetic part of the Lagrangian. 

Explicit expressions of the Lagrangian for the different vertices are
written in Appendix~\ref{alag}. They are derived by substituting the matrices of
Eqs.~(\ref{vm},\ref{psm}) in the generic Lagrangians of 
Eqs.~(\ref{VPP}, \ref{VVP}, \ref{VVV}, \ref{VVVV}, \ref{VVPP}), taking the trace
and rescaling the kaon field as shown in Eq.~(\ref{resc}). From those
Lagrangians it is straightforward to compute the Feynman rules for the vertices. 
We use the standard expressions for the meson propagators 
\be
\la{propk}
D_K (q) = \frac{i}{q^2 - m_K^2 + i\, m_K \Gamma_K^{(tot)} (q^2)}\,, 
\ee
\bea
\la{propks}
D_{K^*}^{\mu \nu} (q) &=&  - \left( g^{\mu \nu} - 
\frac{q^\mu q^\nu}{m_{K^*}^2}\right) \times \nonumber \\[0.2cm]
&& \frac{i}{q^2 - m_{K^*}^2 + i\, m_{K^*}
\Gamma_{K^*}^{(tot)} (q^2)}  \,.
\eea
$\Gamma_{K(K^*)}^{(tot)}$ stand for the total decay widths of the
intermediate $K$, $K^*$ and include all decay channels that are open
for a given $q^2$ value. Using the Lagrangians introduced above one
gets
\bea
\la{widthk}
\Gamma_K^{(tot)} &=& \sum_{a=\rho,\omega,\phi} \left\{ 
\Gamma_{K \raw a \, K}(q^2)\, \theta \left(q^2-(m_a + m_K)^2\right)  
\right.\nonumber \\[0.2cm]
&+& \left.
\Gamma_{K \raw a \, K^*}(q^2) \, \theta \left(q^2-(m_a + m_K^*)^2\right) 
\right\} 
\eea
and
\bea
\la{widthks}
\Gamma_{K^*}^{(tot)} &=& \sum_{a=\pi,\rho,\omega,\phi} \left\{ 
\Gamma_{K^* \raw a \, K}(q^2)\, \theta \left(q^2-(m_a + m_K)^2\right)  
\right.\nonumber \\[0.2cm]
&+& \left.
\Gamma_{K^* \raw a \, K^*}(q^2) \, \theta \left(q^2-(m_a + m_K^*)^2\right) 
\right\}\,,
\eea
where $\theta (x)$ is the standard step function. 
Explicit expressions for the different decay widths are listed in
Appendix~\ref{awidth}. By introducing such imaginary parts in the
propagators, we are implementing an approximate unitarization for the 
s-channel diagrams.  

Since hadrons are extended objects, it
is necessary to insert form factors that suppress high momentum transfers.
This affects all t-channel Feynman diagrams. We adopt the widely used
monopole form 
\be
\la{ff}
F_i(q^2)=\frac{\Lambda^2 - m_i^2}{\Lambda^2 - q^2}\,, \quad i=K,\,K^* 
\ee
assuming the same cutoff parameter $\Lambda=1.8$~GeV for all
species~\cite{Ko:1994id, Haglin:1995xu}. These form factors cause a 
strong reduction in the t-channel contributions to the collision rates. 

The amplitude squared for the VVVV term  is quite large and increases
very fast with $s$ since it involves high powers of both $g$ and $s$ ($g^4$
and $s^4$). As a consequence, higher order corrections become relevant
as one goes away from threshold. We take approximately into account the
resummation of s-channel loops in the contact amplitude for 
$\phi\, +\, a \raw b\, +\, c$   by means of the substitution 
\be
\label{subs}
\bear{l}
\D  \alpha g^2 \raw   \\[0.4cm] 
\D \frac{\alpha g^2}{\left[ 1 + \alpha g^2 G(s,m_\phi,m_a)
\right]^{1/2} \left[ 1 + \alpha g^2 G(s,m_b,m_c) \right]^{1/2}} \,,  
\ear
\ee
with 
\be
\label{G}
G(s,m_a,m_b)= -i \frac{1}{2 \pi} \frac{p_{cm}}{\sqrt{s}} \left[ 1 +
\frac{(s-m_a^2-m_b^2)^2}{8 m_a^2 m_b^2} \right]\,,
\ee
where $p_{cm}$ is the momentum of the vector mesons $a$ and $b$ in the
CM frame. The factor $\alpha = -1/2$ for the $\phi \phi K^* K^*$
vertex and $\alpha = 1/(2 \sqrt{2})$ for $\phi \rho K^* K^*$ and
$\phi \omega K^* K^*$.
This ansatz is inspired by the solution of the
Bethe-Salpeter (BS) equation in the K-matrix approximation using only the part of
the tree-level amplitude, which leads to an algebraic equation (see
Appendix~\ref{B-S} for details). $G(s,m_a,m_b)$ vanishes at threshold, so that
one recovers the tree-level amplitude, and becomes progressively larger with
the increase of $s$, causing an effective suppression of the coupling constant
$g$. At large $s$, our simple expression shall become inaccurate compared with
the solution of the full BS equation with couple-channels
effects incorporated, but one should bare in mind that high values of $s$
are exponentially suppressed in the integral that defines the collision
rates in Eq.~(\ref{CR}). Therefore, this inaccuracy has a small numerical 
impact in the determination of the collision rates, but is important to 
avoid artificially large contributions from cross section values above the
unitarity limit.

With these ingredients, one can
compute the amplitudes; it is important to add them coherently since the
interferences modify appreciably the total cross section. 
A caveat is in order regarding reactions 1.2 and 2.4. In both cases
the exchanged kaon (in the t-channel) can be put on the mass shell, making
the amplitude become singular. The singularity disappears if one takes into
account that kaons will develop an in-medium width~\cite{Ko:1994id, Haglin:1995xu}. 
However, studying how
the modifications of propagators, vertices or masses and widths of the
particles involved in the reactions influence the collision rates goes
beyond the scopes of the present work. Hence, we just do not take them into
consideration. In any case, their contribution will most likely shorten the
mean free path even more.        

\subsection{Results}

The parameters of the HLS model can be fixed using the experimental values
of masses and pseudoscalar decay constants. We choose a=2 after VMD;
then, using the $\rho$ mass in Eq.~(\ref{masses}) and
$f_\pi=92.4$~MeV~\cite{Caso:1998tx} one gets $g=5.89$. Once $a$ and $g$ are
fixed, the remaining part of Eq.~(\ref{masses}) gives
$c_{\chic{V}}$. It ranges from 0.34 to 0.38 depending on weather determined from
$m_{K^*}$ or $m_\phi$. We adopt $c_{\chic{V}}=0.36$. Next, from Eq.~(\ref{fKeta}) and
using the experimental value $f_K/f_\pi=1.22$~\cite{Leutwyler:1984je}, we
obtain $c_{\chic{A}} = 0.49$. Finally, we take $c_{\chic{WZ}}= -0.1$ as determined in
Ref.~\cite{Bramon:1995pq}. Isospin degeneracy is assumed, i.e. the mass
differences between mesons of the same species are not taken into account.                 

The contributions of the different mesons to the $\phi$ collision rate as a
function of temperature are shown in Fig.~\ref{rates}. For temperatures
between 150 and 200~MeV, the collision rate is dominated by the $K^*$ followed
by $K$ and $\rho$, while the contributions from $\pi$, $\omega$ and $\phi$
are smaller.  
\bfig[h!]
\bcen
\includegraphics[width=0.9 \linewidth]{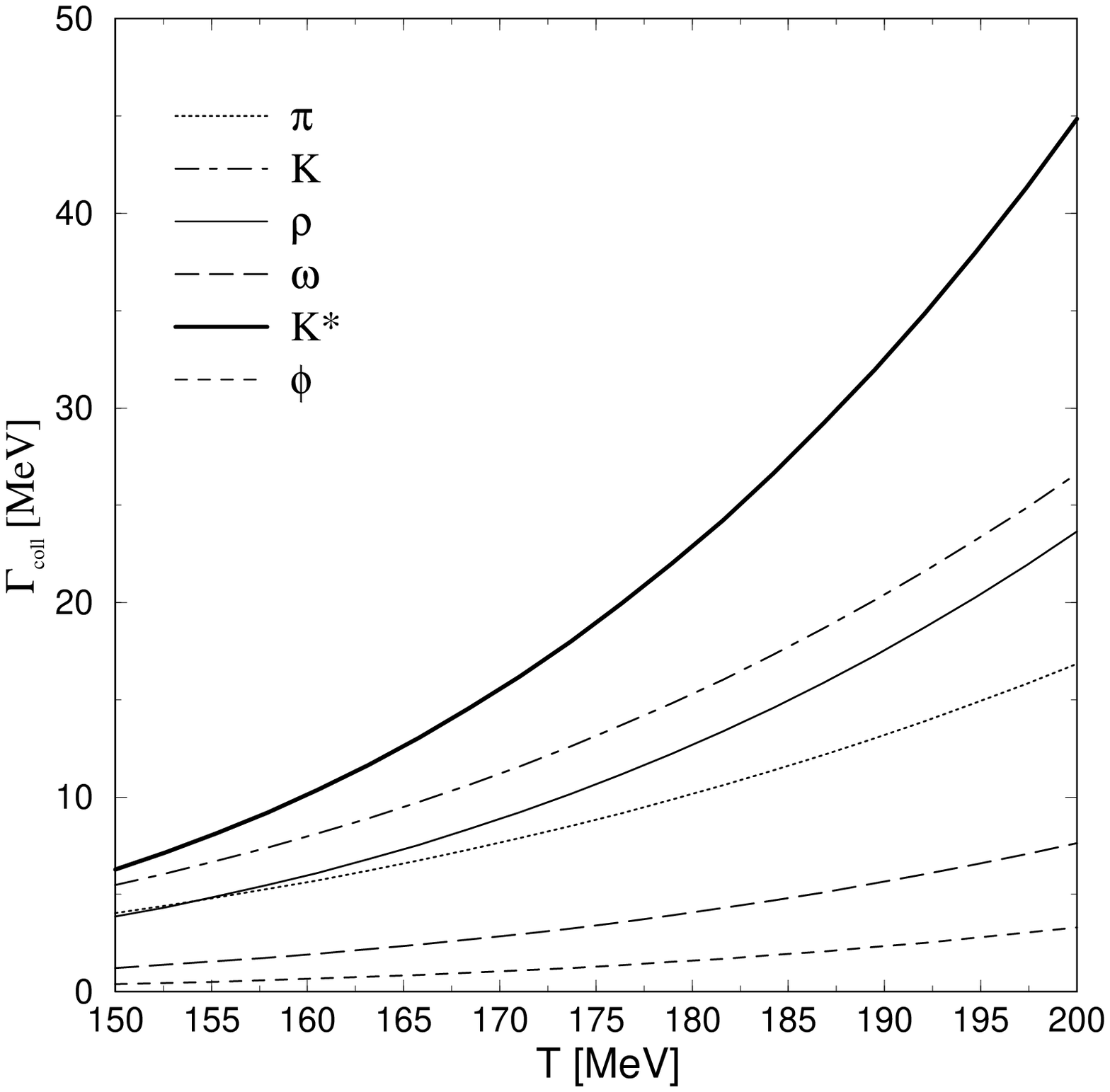}
\caption{Collision rates of $\phi$ with $\pi$, $K$, $\rho$, $\omega$, $K^*$
and $\phi$ as a function of temperature.}
\la{rates}
\ecen
\efig

In order to understand why $\Gamma_{coll}^{(K^*)}$ is bigger than
$\Gamma_{coll}^{(K)}$ in spite of their mass difference
let us recall that $\Gamma_{coll}^{(a)}$ can be cast as 
\be
\la{CResq} 
\Gamma_{coll}^{(a)} = n_a \langle \sigma v_{rel} \rangle \,.
\ee
Hence, the averaged rate factor $\langle \sigma v_{rel} \rangle$ must
overcome a ratio of  (see Eq.~\ref{den}) 
\be
\la{KKs}
\frac{n_{K^*}}{n_K}=\frac{3 m_{K^*}^2 K_2(m_{K^*}/T)}{m_K^2
K_2(m_K/T)}\,;
\ee
at $T=200$~MeV this ratio is equal to 0.77. This number is much bigger than
what one would naively expect. In fact, the small number coming from the
ratio of Bessel functions is almost compensated by the ratio of squared masses
times the spin degeneracy. The former being due to the higher density of 
states for a heavy particle. Therefore, a larger $\Gamma_{coll}^{(K^*)}$ can
be achieved by a {\it moderately} larger cross section. The comparison between
Figs.~\ref{Ksrates} and \ref{Krates} show that the
reactions involving all four vector mesons make the true difference between 
$\Gamma_{coll}^{(K^*)}$ and $\Gamma_{coll}^{(K)}$, and the mechanisms
involving VVVV and VVV vertices given in Eqs.~(\ref{VVV}, \ref{VVVV}) are
responsible for this. 
\bfig[h!]
\bcen
\includegraphics[width=0.9 \linewidth]{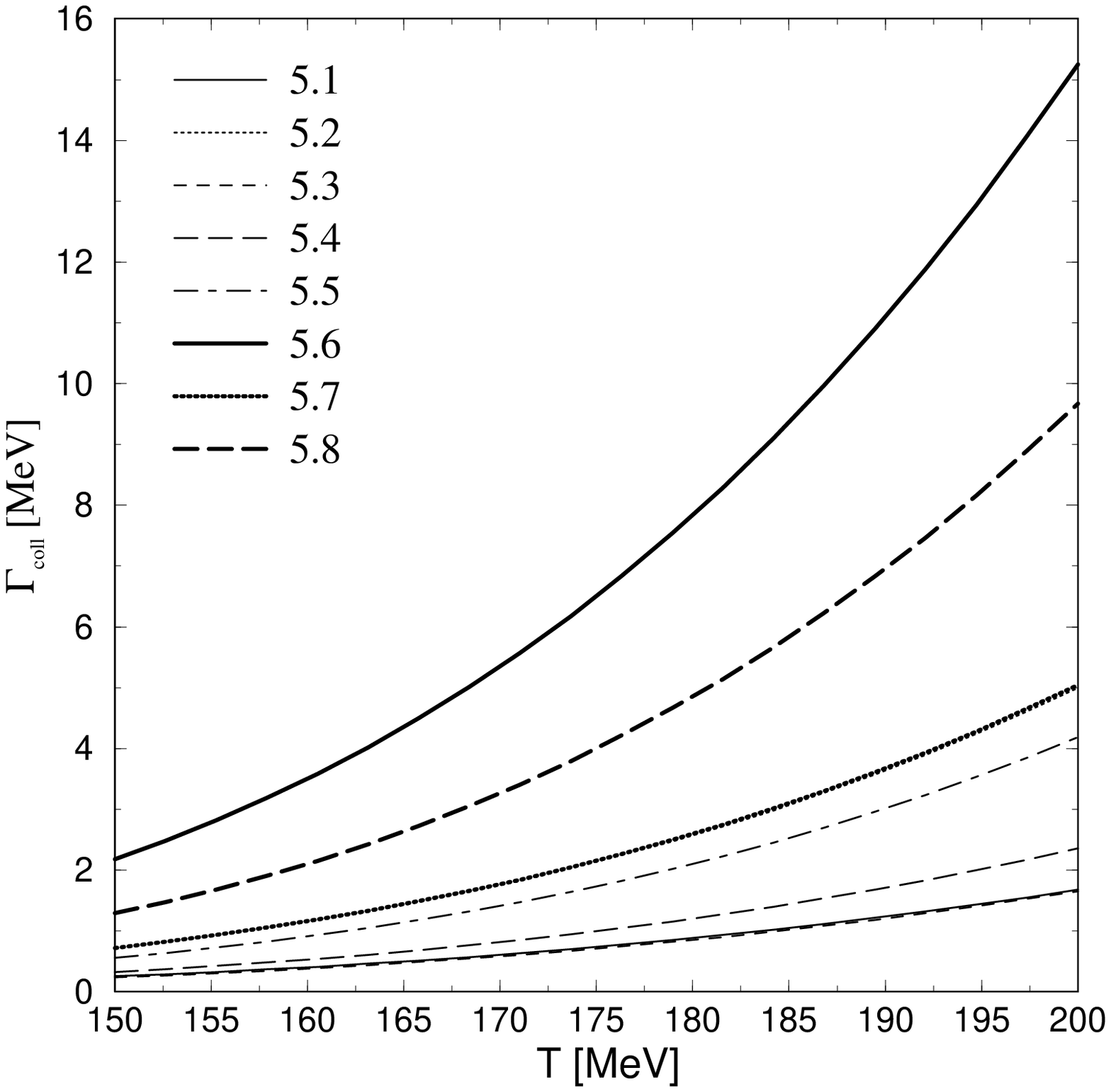}
\caption{Contribution from the various reactions of $\phi$ with $K^*$
to the collision rates. The labels correspond to the reaction numbers in
Table~\ref{table}. The largest rate is from 
$\phi\, +\, K^* \raw \rho\, +\, K^*$. Notice that the contributions of
reactions 5.1 and 5.3 as well as 5.2 and 5.7 almost coincide and are hardly
distinguishable in the plots.}
\la{Ksrates}
\ecen
\efig
\bfig[h!]
\bcen
\includegraphics[width=0.9 \linewidth]{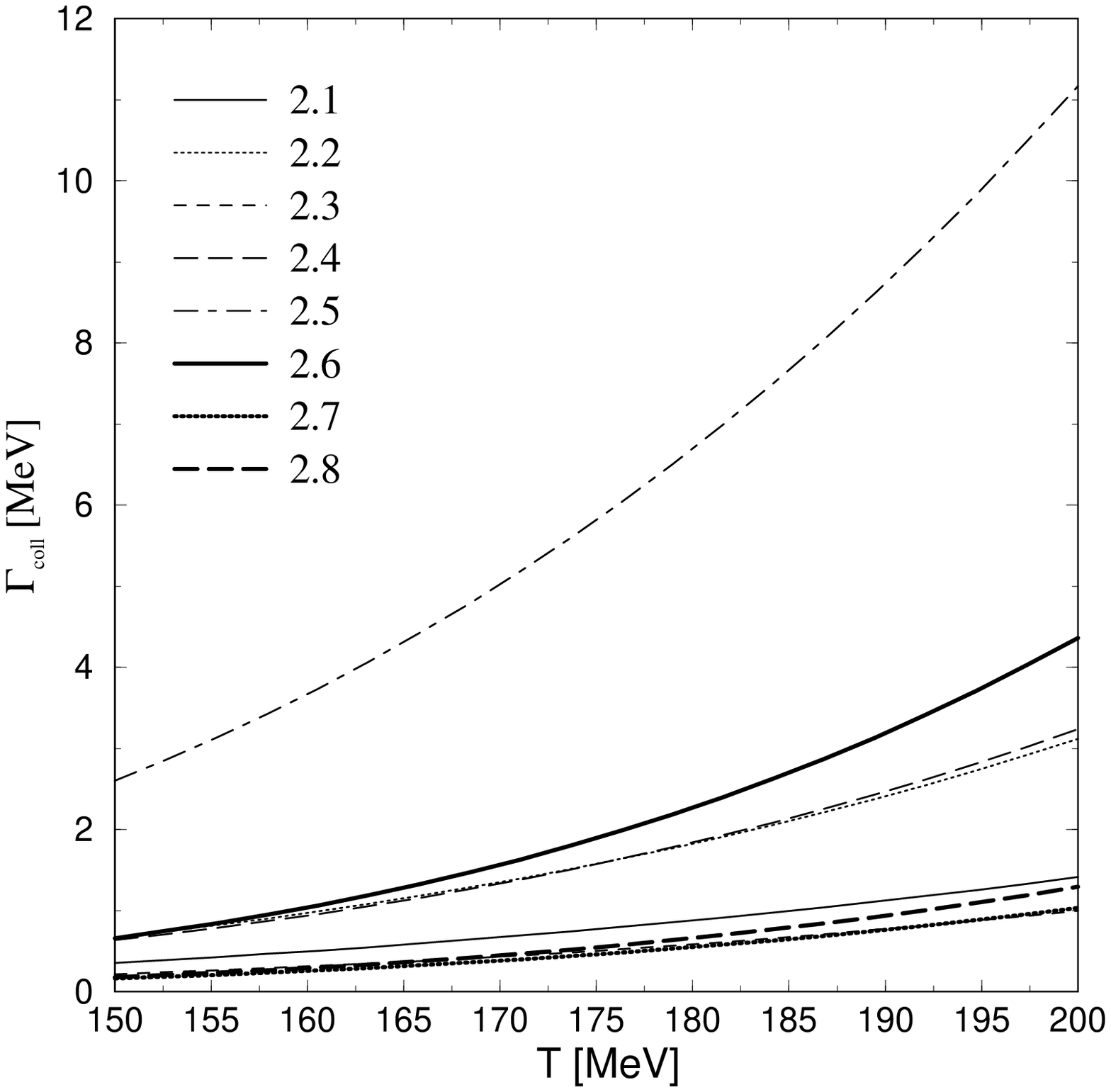}
\caption{Contribution to the collision rate of $\phi$ with $K$ from the
different reactions listed in Table~\ref{table}.}
\la{Krates}
\ecen
\efig

The sum of all partial rates of Fig.~\ref{rates} is shown in
Fig.~\ref{Total}. It is considerably bigger than what has been predicted in
all previous works~\cite{Ko:1994id, Haglin:1995xu, Smith:1998xu}. In this
case, the major novel ingredient is the dynamics involving different vector
mesons as shown above. The dashed line represents the contribution of all
$\phi$-number changing reactions; that is, all reactions in
Table~\ref{table} except the (quasi)elastic ones 2.4, 2.8, 5.4 and 5.8. 
The inelastic reactions account for more than 80~\% of the total collision
rate. 
\bfig[h!]
\bcen
\includegraphics[width=0.9 \linewidth]{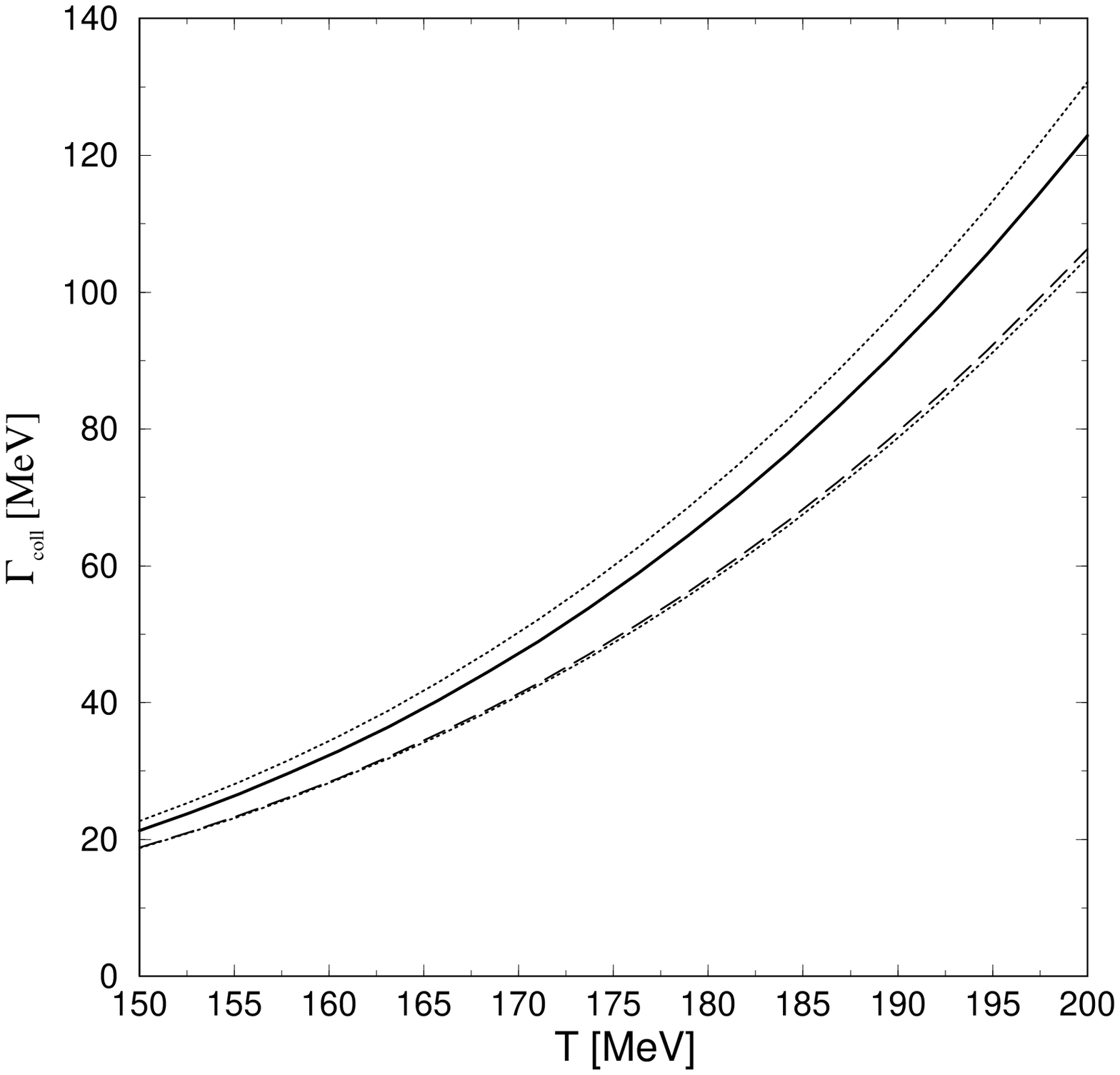}
\caption{Total collision rate of $\phi$ meson as a function of temperature
(solid line). The dotted lines constrain the region of possible values when the
model parameters are changed as described in the text. The dashed line is
the contribution from $\phi$-number changing processes alone.}
\la{Total}
\ecen
\efig

We have also studied the sensitivity of the result to the modification of 
parameters in the HLS model. It was found to depend appreciably on the value
of $g$, which is not surprising since the cross sections involve high powers
of this coupling (up to $g^8$). As mentioned in Section~\ref{HLS}, the value
of $a$ extracted from the experiment is slightly bigger than 2. If we
take $a=2.4$~\cite{Benayoun:1998ex}, then a lower limit of $g=5.38$ is obtained.
We fix the upper limit by using  $m_\omega$ instead of
$m_\rho$, getting $g=5.98$. The region of possible values of $\Gamma_{coll}$
is bounded by the dotted lines in Fig.~\ref{Total}.

In Fig.~\ref{mfpath} we show the $\phi$ mean free path as a function of 
temperature. It goes below 2.4~fm at $T > 170$~MeV. This is much 
smaller than the typical size of the hadronic system created in
heavy ion collisions (10-15~fm). Therefore, contrary to the common
believe~\cite{Shor:1985ui}, the $\phi$ mesons that are created after
hadronization, at temperatures presumably between 170 and 200~MeV, will not
leave the fireball without interacting; they will most likely be absorbed
and reemitted in the hadronic phase. 
The uncertainties discussed above do not modify this conclusion.         
\bfig[h!]
\bcen
\includegraphics[width=0.9 \linewidth]{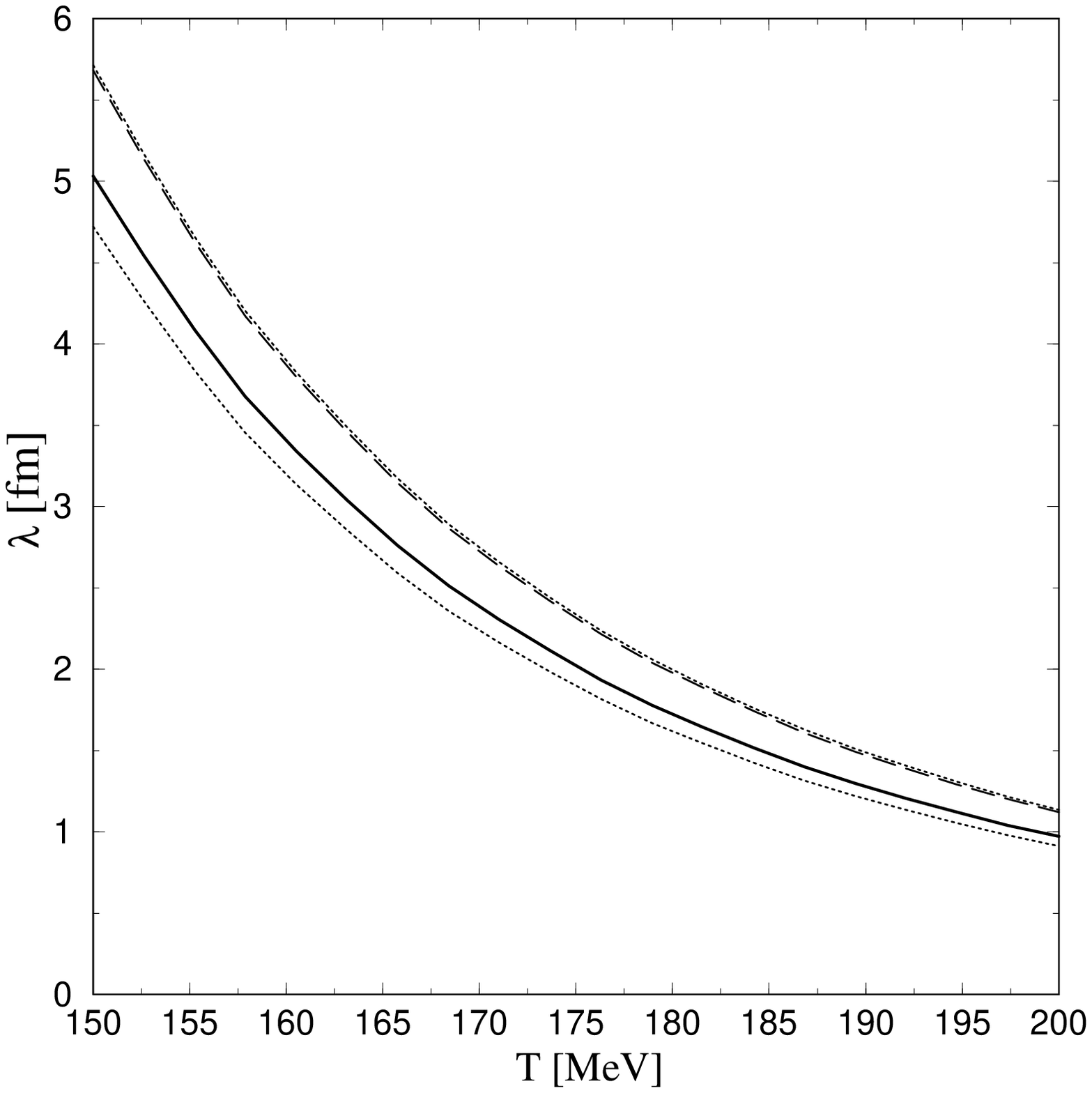}
\caption{Temperature dependence of the $\phi$ mean free path in hot 
hadronic matter. Line styles have the same meaning as in Fig.~\ref{Total}.}
\la{mfpath}
\ecen
\efig

\section{Time evolution} 

We now proceed to study qualitatively the time evolution of the $\phi$
density in the expanding hadronic fireball. It is governed by the following
rate equation
\be
\la{rateq}
\partial_\mu \left( n_\phi u^\mu \right) = \Psi
\ee
where $u^\mu=\gamma (1,\bnu)$ is the four velocity, defined in terms of the
Lorentz factor $\gamma$ and the fluid velocity $\bnu$; $\Psi$ stands for the
source term 
\bea
\la{source}
\Psi = &-& \sum_{a,b,c} \langle \sigma_{\phi a \raw b c} v_{\phi a} 
\rangle n_\phi n_a + \sum_{a,b,c} \langle \sigma_{b c \raw \phi a} 
v_{b c} \rangle n_b n_c\nonumber \\[0.2cm]   
&-& \sum_{b,c} \langle \Gamma_{\phi \raw b c} \rangle n_\phi  + 
\sum_{b,c} \langle \sigma_{b c \raw \phi} v_{b c} \rangle n_b n_c \,,
\eea 
which takes into account $\phi$-number changing processes: binary
collisions, decay, recombination. 
In this approach, the evolution of momentum distribution is not
investigated. Kinetic equilibrium (and the applicability of the
hydrodynamical description) is assumed, but not chemical 
equilibrium~\cite{Biro:1983yo,Ko:1988mf}. 

In principle, Eq.~(\ref{rateq}) is just one in a system of coupled equations
that describes the time evolution of the densities of all species. Such a
treatment would require the knowledge of the reaction cross sections
for all the mesons of the model. We shall rather perform the simplifying
assumption that, at hadronization temperature, all particles 
($\pi$, $K$, $\rho$, $\omega$, $K^*$, $\phi$) are in chemical equilibrium, 
and all of them except the $\phi$ remain in it until freezeout. In chemical
equilibrium, detailed balance holds, i.e.  
\be
\la{detb1}
\langle \sigma_{\phi a \raw b c} v_{\phi a} \rangle n_\phi^{eq} n_a^{eq} = 
\langle \sigma_{b c \raw \phi a} v_{b c} \rangle n_b^{eq} n_c^{eq}
\ee
and
\be
\la{detb2}
\langle \Gamma_{\phi \raw b c} \rangle n_\phi^{eq} = 
\langle \sigma_{b c \raw \phi} v_{b c} \rangle n_b^{eq} n_c^{eq}\,.
\ee
Substituting these equations in Eq.~(\ref{source}) and imposing that 
particle number densities for all species except $\phi$'s take their
equilibrium values: 
\be
\la{asum}
n_{a,b,c} = n^{eq}_{a,b,c}\,,
\ee 
one obtains 
\be
\la{rateqsim}
\partial_\mu \left( n_\phi u^\mu \right) = - \Gamma \left( n_\phi - n_\phi^{eq}
\right) \,,
\ee
where $n_\phi^{eq} (T)$ comes directly from Eqs.~(\ref{den}, \ref{bolz})
\be
\la{phideq}
n_\phi^{eq} (T) = \frac{3}{2 \pi^2} m_\phi^2 T K_2\left(\frac{m_\phi}{T} \right) \,;
\ee
$\Gamma (T)$ denotes the number of interactions per unit time and   
is split in two parts 
\be
\la{cpd}
\Gamma (T) = \Gamma_{coll} (T) + \Gamma_{dec} (T)\,.
\ee

$\Gamma_{coll}$ is the collision rate calculated in the previous Section.
Notice however that the right hand side of Eq.~(\ref{rateqsim}) does not
hold for collisions of two $\phi$ mesons
$\phi\, +\, \phi \raw b\, +\, c$ (reactions 6.1-6.3). The reason being
that $\phi$'s are not in chemical equilibrium. For our simple estimate 
we thus ignore those processes which are very small (see short-dashed line in 
Fig~\ref{rates}). Analogously, the contribution
of (quasi)elastic reactions $\phi\, +\, a \raw \phi\, +\, c$
(2.4, 2.8, 5.4, 5.8) to the source term $\Psi$ vanishes if particles 
$a$ and $c$ are in chemical equilibrium. As we have already stated, these
reactions account for less than 20~\% of the total collision rate in 
Fig~\ref{Total}.  

$\Gamma_{dec}(T)$ is the
average free decay width. For the sake of consistency we only consider
decays into kaon pairs, which account for 83~\% of the total width. The
general expression for $\Gamma_{dec}$ is a trivial modification of
Eq.~(\ref{CRgen}). Proceeding as in Section~\ref{general} one gets
\be
\la{fwidth}
\Gamma_{dec} (T) = \Gamma_0 \frac{K_1 (m_\phi/T)}{K_2 (m_\phi/T)}\,,
\ee
where $\Gamma_0 = 3.7$~MeV. Obviously, at high temperatures, $\Gamma_{dec}$
is negligible compared to $\Gamma_{coll}$, but the later drops fast when the
system cools down.    

In central relativistic heavy ion collisions the distribution of matter is
approximately uniform in rapidity, at least in the central rapidity region,
and the geometry of the collision is cylindrically
symmetric~\cite{Bjorken:1983qr}.  Therefore, it
is convenient to adopt as coordinates $\{ \tau , \eta, r , \varphi \}$ where
$r$ and $\varphi$ are the transverse radius and polar angle, while $\tau$ and
$\eta$ represent the {\it longitudinal} proper time and space-time rapidity
respectively
\be
\la{variab}
\tau = \sqrt{t^2 - z^2}\,, \quad \eta = \frac{1}{2} \ln{\frac{t+z}{t-z}}\,.
\ee
The assumptions of a radial transverse expansion and  Lorentz invariance in
the central region imply that $u^\eta = u^\varphi = 0$. Next, assuming a
uniform density distribution in the transverse plane and averaging over the
radial coordinate (analogously to what was done in
Refs.~\cite{Biro:1983yo,Ko:1988mf} for the spherically symmetric case) we get 
\be
\la{rateqf} 
\frac{1}{\tau R^2(\tau)} \frac{\partial}{\partial \tau} \left\{ \tau
R^2(\tau) n_\phi \langle u^\tau \rangle \right\} = -\Gamma (T) \left( n_\phi
-n_\phi^{eq} \right) \,,  
\ee
which is to be solved with the initial condition $n_\phi(\tau_0) = n_\phi^{eq}
(T_0)$, $\tau_0$ and $T_0$ being the hadronization time and temperature;
$R(\tau)$ is the transverse radius of the system. Notice that $\tau \pi
R^2$ is nothing but the volume of the expanding fireball at mid-rapidity. 
The averaged $\tau$ component of the four velocity is given by 
\be
\la{aveu}
\langle u^\tau \rangle = \frac{2}{R^2 (\tau)} \int_0^{R(\tau)} dr\, r u^\tau
(r)\,.
\ee

Following Ref.~\cite{Schnedermann:1993hp}, we assume that the flow vector
$u^\mu$ can be constructed from two independent boosts, one in the longitudinal
and another in the radial direction.  Then, it is easy to see that at
mid-rapidity  
\be
\la{ut}
u^\tau = \gamma_r = \frac{1}{\sqrt{1 - \beta_r^2}} \,.
\ee
A reasonable ansatz for radial velocity is~\cite{Schnedermann:1993hp} 
\be
\la{anz} 
\beta_r (\tau,r)  = \beta_s \left( \frac{r}{R} \right)^a\,. 
\ee
In the case of a constant $\beta_s$
\be
\la{uave} 
\langle u^\tau \rangle = \int_0^1 dy\, \frac{1}{\sqrt{1 - \beta_s^2 y^a}}
\ee
is a constant too. Using the boundary condition 
\be
\la{bound}
\frac{d R}{d \tau} =  \beta_r (\tau,R)=\beta_s
\ee
one gets 
\be
\la{Rt}
R(\tau) = \beta_s (\tau - \tau_0 ) + R_0 \,,
\ee
where $R_0 = 1.2\, A^{1/3}$~fm is the radius of the colliding nuclei.   

Finally, in order to solve the rate equation we need the relation between
time and temperature, which can be derived from entropy conservation 
\be
\la{ent}
\partial_\mu \left( s u^\mu \right) = 0\,.
\ee
Repeating the same procedure as for $n_\phi$ one finds  
\be
\la{avent}
\frac{1}{\tau R^2(\tau)} \frac{\partial}{\partial \tau} \left\{ \tau
R^2(\tau) n_\phi \langle u^\tau \rangle \right\} =0\,.
\ee
This equation can be easily solved, obtaining the following implicit
expression for $\tau(T)$ (or vice versa)
\be
\la{tauT}
\tau R^2(\tau) \langle u^\tau \rangle(\tau)  s(T) =\tau_0 R_0^2 \langle
u^\tau \rangle(\tau_0)  s(T_0)\,.
\ee
If $\langle u^\tau \rangle$ does not depend on $\tau$, as for the ansatz
considered above, Eq.~(\ref{tauT}) reflects that the total entropy 
in the central region of the collision is conserved.

The functional dependence of the entropy  upon the temperature depends on
the properties of the system under consideration. We describe it as an ideal
gas of $\pi$, $K$, $\rho$, $\omega$ and $K^*$ obeying Boltzmann statistics
\be
\la{gas}
s(T) = \sum_i \frac{g_i}{2 \pi^2} m_i^3  K_3\left( \frac{m_i}{T} \right)\,,
\ee
with $i=\{ \pi, K, \rho, \omega, K^* \}$; g denotes the degeneracy (both spin
and isospin). The underlying idea is that the equation of state of a gas of 
free hadrons and resonances should mimic the one of a dense hadron
gas~\cite{Sollfrank:1997hd}.  

Fig.~\ref{mres} shows the $\tau$ dependence of the ratio of the
$\phi$ yield at mid-rapidity $N(\tau) = \tau \pi R^2(\tau) n_\phi(\tau)$ and
the number of them right after hadronization
$N_0 = \tau_0 \pi R_0^2 n^{eq}_\phi (T_0)$. We have used $\tau_0= 1$~fm, $T_0 =
190$~MeV, $\beta_s = 0.6$, $a=1$ and $A=197$~(Au).           
\bfig[h!]
\bcen
\includegraphics[width=0.9 \linewidth]{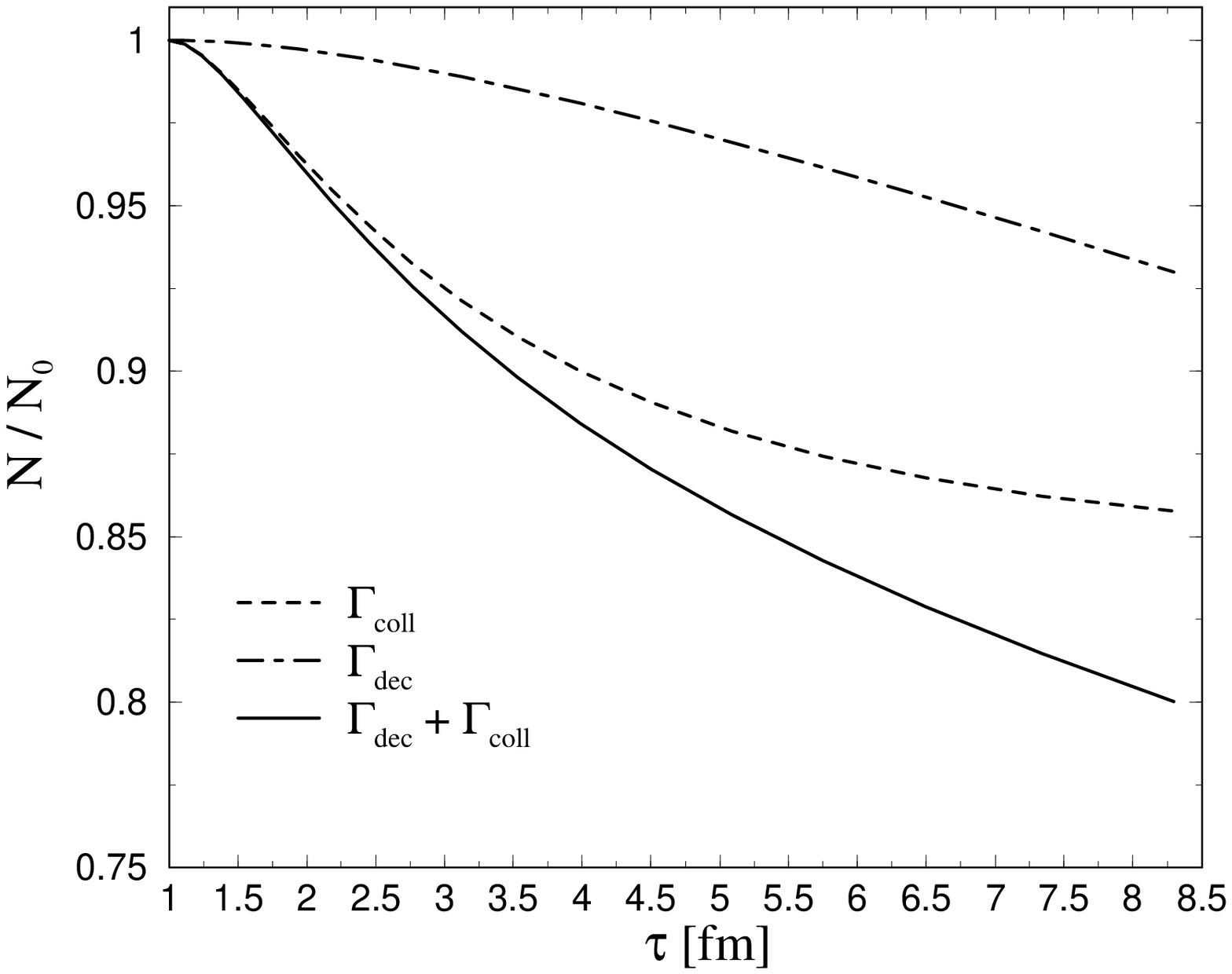}
\caption{Time dependence of the ratio of the $\phi$ number to its value at 
hadronization. The dashed line shows the contribution from inelastic
collisions while the long-short dashed one accounts for free decay (and
recombination) The combined effect of both collisions and free decay is 
described by the solid line. }
\la{mres}
\ecen
\efig
 
At the early stages of the expansion, the system is hot and the collision
rates are high, so that $\phi$'s are pushed towards equilibrium making its
number decrease fast. As the system cools down, the collision rates become
small and the contribution of the collision rates to the ratio starts to
saturate. The effect of the decay is negligible below $\tau =2$~fm, not only
because $\Gamma_{dec} \ll \Gamma_{coll}$ but also because detailed
balance works more efficiently than at lower temperatures. We have taken a
freezeout temperature of $T_f = 100$~MeV, which corresponds to a life time of
$\tau_f = 8.3$~fm. If one assumes that only those $\phi$ mesons present at 
freezeout are detectable via kaon pairs, then the $\phi$ yield is reduced in
20~\% with respect to hadronization.    

This model exhibits a notable dependence on the value of $\beta_s$, as
illustrated in Fig~\ref{betasdep}. The bigger $\beta_s$, the faster the
temperature drops and, hence, the closer $N/N_0$ gets to one.    
\bfig[h!]
\bcen
\includegraphics[width=0.9 \linewidth]{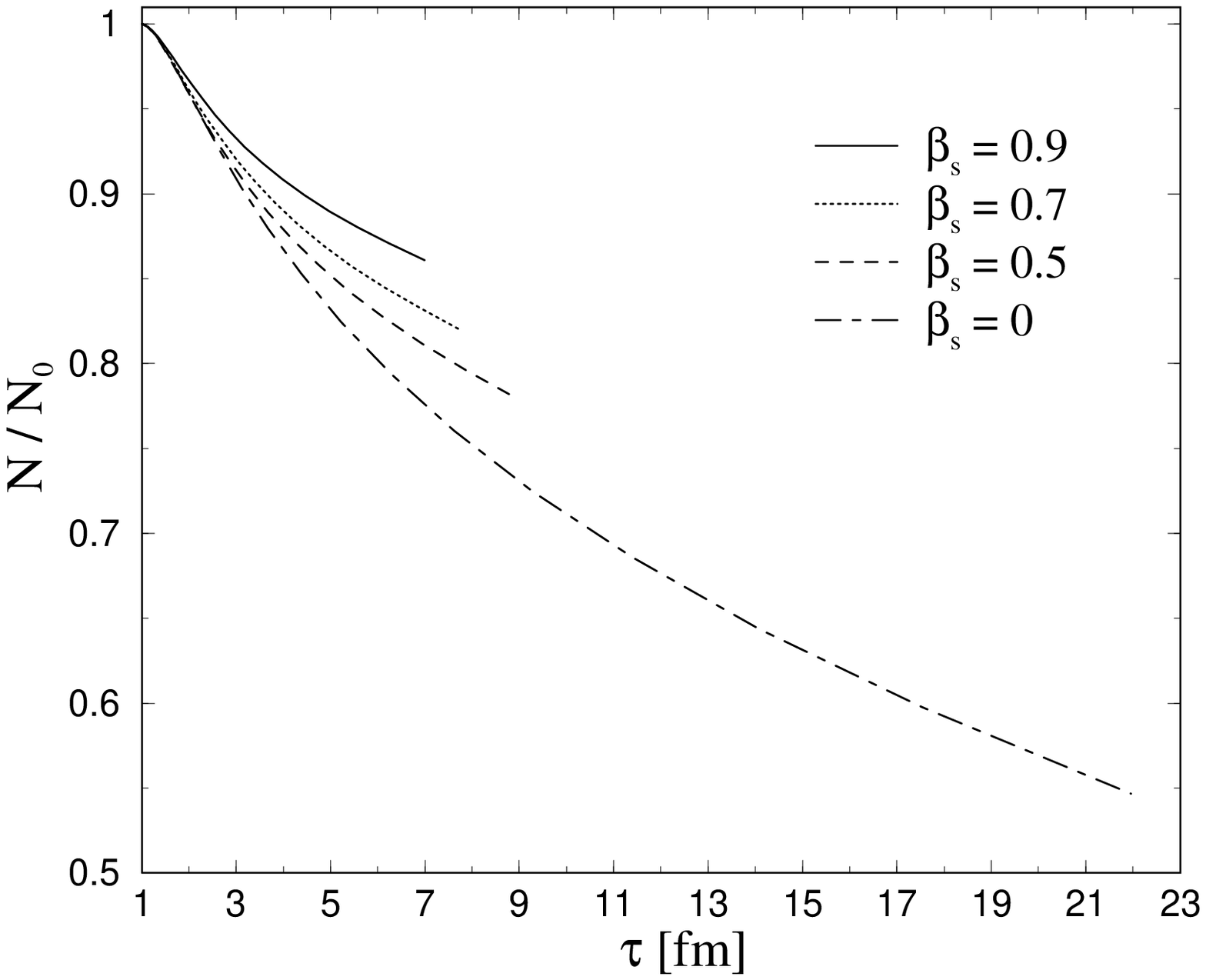}
\caption{Ratio of the $\phi$ number to its value at hadronization for
different values of the transverse flow velocity at the surface. The end
point of the lines correspond to a temperature of 100~MeV.}
\la{betasdep}
\ecen
\efig
The case $\beta_s = 0$ corresponds to the Bjorken limit, where the
transverse expansion is neglected. For our choice of the equation of state,
this is a rather unrealistic situation because the system lives so long that
the transverse expansion can not be neglected. For more realistic values of
$\beta_s$ we get a ratio $N/N_0$, at $T_f = 100$~MeV,  between 0.78 and 0.86. 

We have also considered different hadronization temperatures, as shown in
Fig.~\ref{T0dep}. If the hadronization temperature is low, the collision
rates are small from the very beginning, and the interactions become less
relevant.   
\bfig[h!]
\bcen
\includegraphics[width=0.9 \linewidth]{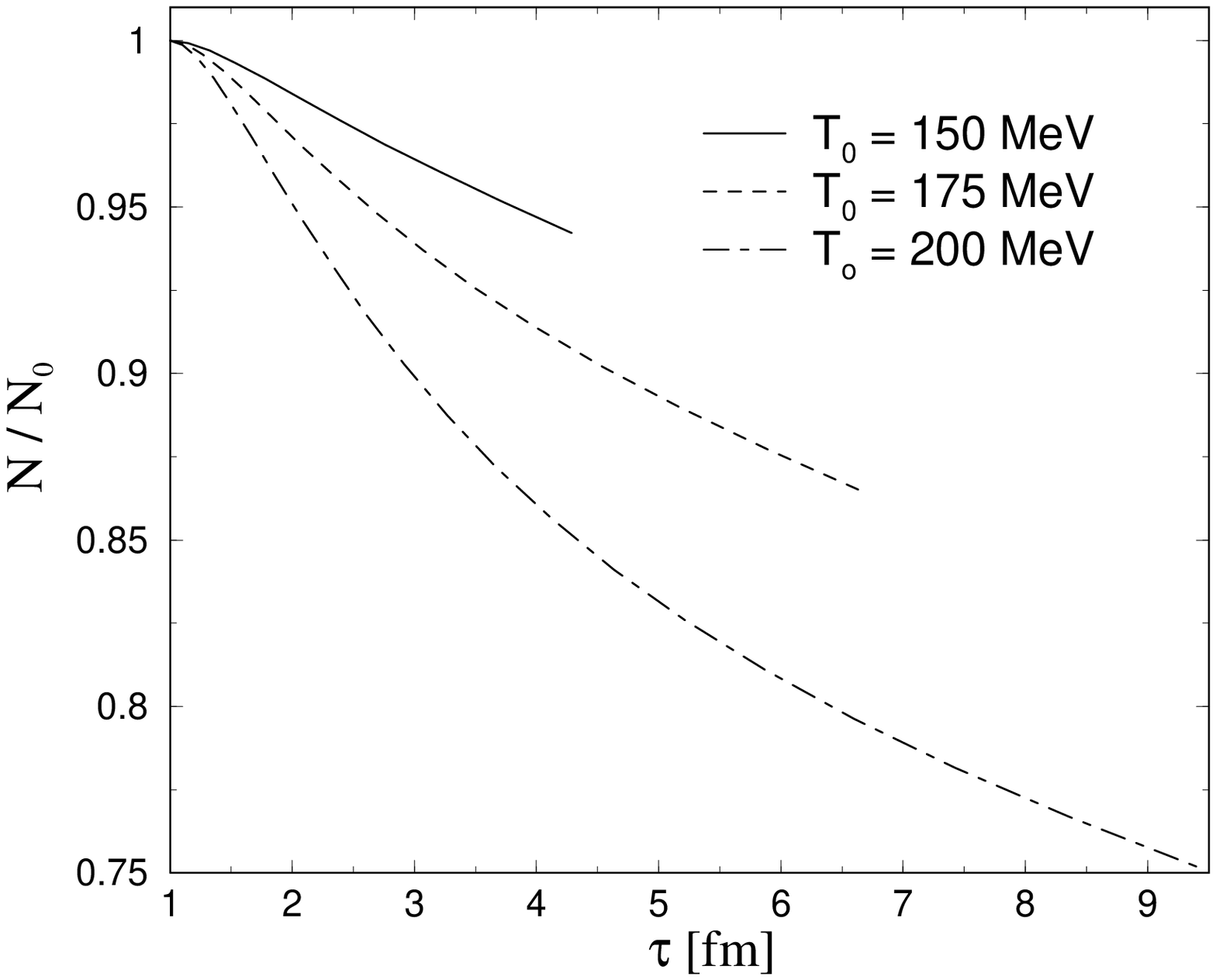}
\caption{Same as Fig.~\ref{betasdep} but for different hadronization
temperatures.}
\la{T0dep}
\ecen
\efig

The present simple  description of the time evolution of the $\phi$ number
in an expanding hadronic gas is intended to illustrate the consequences of
the short $\phi$ mean free path that we have obtained. A realistic
analysis of the spectra would require a more sophisticated treatment.
In particular, the fact that the $\phi$ collision rate is dominated by
$\phi$-number changing reactions implies that mainly those processes are
responsible for maintaining the kinetic equilibrium via detailed balance.
Therefore, if the system is out of chemical equilibrium, detailed balance no
longer holds and the system will also depart from kinetic equilibrium.
This means that both $\phi$ temperature and flow velocity will differ from
those of other species with high rates of elastic scatterings like pions. 

\subsection{The role of chemical potentials} 

Up to now we have not introduced chemical potentials (fugacity factors) for
any of the particle species under consideration. However, it is well known
that when the number changing processes are not effective, particle numbers
are fixed and chemical potentials associated with these conserved quantities
appear~\cite{Bebie:1992ij,Song:1997ik}. For pions and kaons values of
$\mu_{\pi} = 60-80$~MeV and $\mu_K = 100-130$~MeV are reached at freezeout temperatures
between 110 and 120~MeV~\cite{Rapp:2000gy}.  

In order to study the effect of such conserved meson numbers in the $\phi$
yield we assume that
$\mu_i = 0$ at hadronization for all species and grows to a value $\mu^f_i$
at freezeout. A simple linear interpolation is used in between: 
\be
\la{interp}
\mu_i(T)=\frac{T_0 - T}{T_0 - T_f} \mu^f_i \,.
\ee
We take $\mu^f_{\pi} = 70$~MeV, $\mu^f_K = 115$~MeV with $T_f =115$~MeV
to be consistent with the values quoted above. Since
$\rho \leftrightarrow \pi \pi$ is a fast process, one can say that $\pi$ and
$\rho$ mesons are in relative chemical equilibrium i.e.
$\mu_\rho = 2 \mu_\pi$~\cite{Song:1997ik}. In the case of the $\omega$,
a recent calculation has obtained a large collision rate for the reaction
$\omega \pi \raw \pi \pi$ at high temperatures~\cite{Schneider:2001zt};
therefore, we take $\mu_\omega = \mu_\pi$. With the $K^*$ the situation is
uncertain because its decay width into $K \pi$ (51~MeV) might or might not
be large enough to ensure a relative chemical equilibrium. Since the $K^*$'s
are important for the $\phi$ yield due to their large contribution to the
collision rate, we consider here two possibilities. The first is the
assumption of chemical equilibrium  via
$K^* \leftrightarrow K \pi$, so that $\mu_{K^*} = \mu_K + \mu_{\pi}$ and the
second is that the $K^*$ has its own chemical potential, which we keep equal
to the initial one $\mu_{K^*} = 0$ for the sake of simplicity.

The presence of chemical potentials implies that the condition (\ref{asum}) 
should be replaced by 
\be
\la{chemas}
n_{a,b,c} = n^{\mu}_{a,b,c} = n^{eq}_{a,b,c} e^{\frac{\mu_{a,b,c}(T)}{T}}\,,
\ee
leading to the following rate equation 
\bea
\la{ratechem}
\partial_\mu \left( n_\phi u^\mu \right) &\! =\! \!\!&-\sum_{a,b,c} 
\Gamma_{\phi a \raw bc} \left( n_\phi e^{\frac{\mu_a(T)}{T}}  - n_\phi^{eq} 
e^{\frac{\mu_b(T)+\mu_c(T)}{T}} \right) \nonumber \\[0.4cm]
&&-\Gamma_{dec} \left( n_\phi - n_\phi^{eq} 
e^{\frac{2\mu_K(T)}{T}} \right)\,. 
\eea
The equation of state should also be modified. Now, instead of 
Eq.~(\ref{gas}), we have 
\bea
\la{gaschem}
s(T) &=& \sum_i \frac{g_i}{2 \pi^2} m_i^3  \left\{ K_3\left( \frac{m_i}{T} 
\right) \right. \nonumber \\[0.2cm] 
&-& \left. \left[ \frac{\mu_i (T)}{m_i} - \frac{T}{m_i} 
\frac{d\mu_i (T)}{d T} \right] 
K_2\left( \frac{m_i}{T}\right)   
\right\}\,. 
\eea

The result for the ratio $N/N_0$ is given in Fig.~\ref{Chemdep}. 
\bfig[h!]
\bcen
\includegraphics[width=0.9 \linewidth]{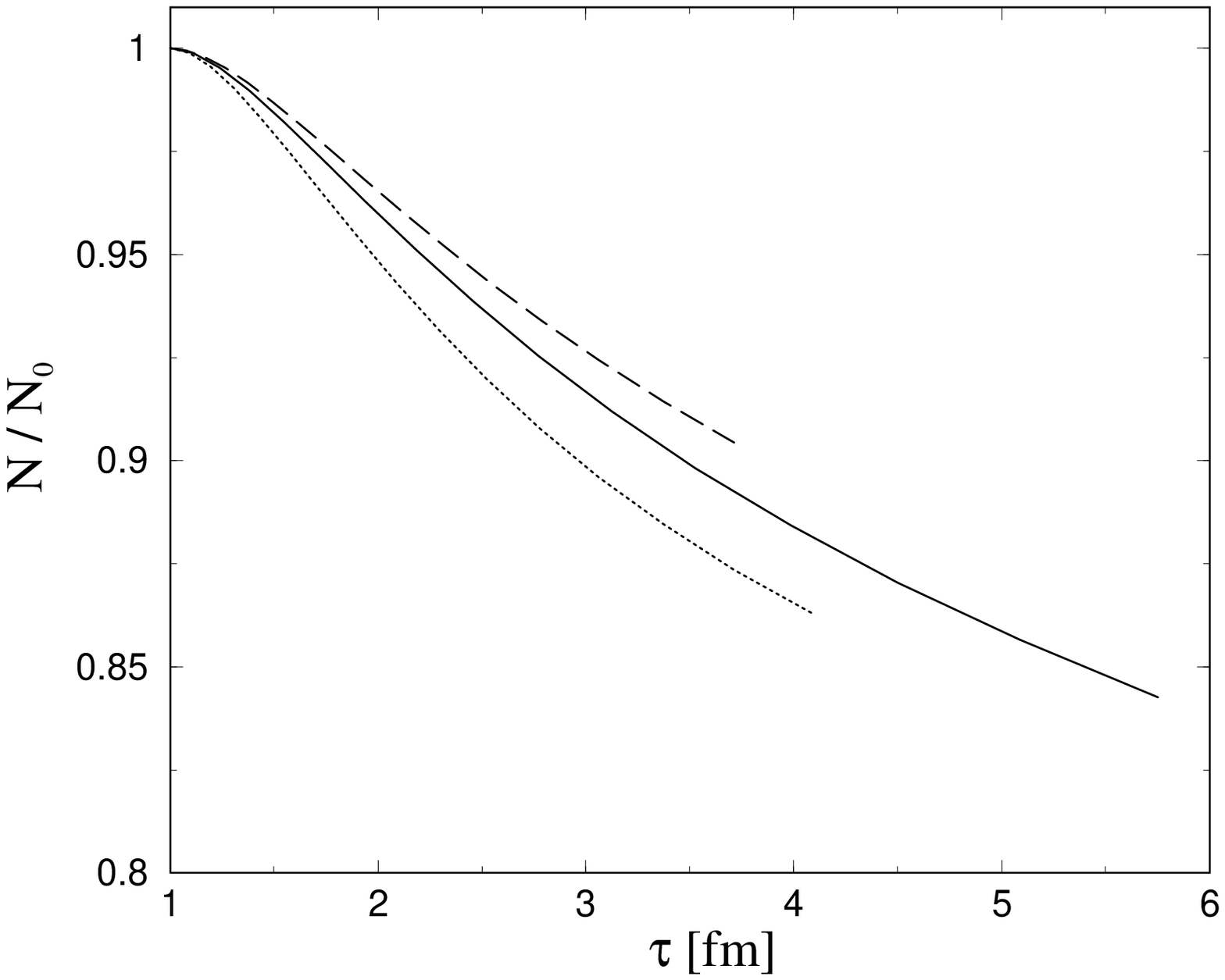}
\caption{Time dependence of the ratio $N/N_0$ at zero (solid line) 
and nonzero (dashed line) chemical potentials, as described in the text. 
The dotted line stands for the case of $\mu_{K^*} = 0$ while keeping 
$\mu_{\pi,K,\rho,\omega} \neq 0$.}
\la{Chemdep}
\ecen
\efig
The solid line reproduces the calculation without chemical potentials. It 
corresponds to the solid line in Fig.~\ref{mres} but ends earlier due to the 
higher freezeout temperature considered here. The introduction of chemical 
potentials causes an increase in the number of $\phi$'s as shown by the 
dashed line. In this case, the inverse reactions are favored keeping the 
$\phi$ number higher. The higher density  at a given temperature also 
translates into a higher pressure and, therefore, a faster cooling and 
earlier freezeout.     
If one sets $\mu_{K^*}$ to zero while keeping the previous values 
for the other mesons (dotted line) then 
the tendency is inverted, and 
direct reactions are the most efficient. However, the final $\phi$ number is 
still bigger than in the $\mu_i = 0$ case due to the faster cooling of the 
hadron gas.   
          
\section{Summary}  

We have studied $\phi$ interactions with a hot hadronic medium composed of
$\pi$, $K$, $\rho$, $\omega$, $K^*$ and $\phi$ using the Hidden Local
Symmetry model. In this way, we could take into account many vertices that are
allowed by the symmetries of strong interactions but whose couplings cannot
be directly determined experimentally. This is the case of three and
four vector meson vertices which are responsible of the large collision
rates of $\phi$'s with other vector mesons, specially $K^*$. As a
consequence, we have shown that the $\phi$ mean free path at temperatures
above 170~MeV is between 2.4 and 1~fm i.e. much smaller than the typical size of
the hadronic system created in relativistic heavy ion collisions. 

The implications of this result for the $\phi$ yield  has been investigated
by solving the rate equation for the $\phi$ density assuming kinetic but not
chemical equilibrium. The high collision rates at the early stages of the
hadronic evolution tend to maintain the equilibrium, causing a reduction of
the $\phi$ number with respect to hadronization. This decrease ranges from
5~\% to 45~\% depending on hadronization temperature, freezeout
temperature and flow velocity. Finally, we have studied how the
departure from chemical equilibrium of $\pi$, $\rho$, $K$ and $K^*$ mesons, 
taken into account the by introducing chemical potentials, influences the time 
evolution of the $\phi$ number.

\begin{acknowledgments}
We thank P. Huovinen and N. Xu for many helpful discussions. L.A.R.
acknowledges a fellowship from the Spanish Ministry of Education and Culture.
He also acknowledges the hospitality at the Theory Department, GSI, where
part of this work was performed. 
This work was supported by the Director,
Office of Science, Office of High Energy and Nuclear Physics,
Division of Nuclear Physics, and by the Office of Basic Energy
Sciences, Division of Nuclear Sciences, of the U.S. Department of Energy
under Contract No. DE-AC03-76SF00098.
L.A.R. has also been supported in part by Spanish DGICYT contract 
number BFM2000-1326. 
\end{acknowledgments}

\appendix
\section{Full set of  Lagrangians}
\la{alag}

The explicit expressions of the Lagrangians describing the interaction
vertices used in this work are listed below. 

\begin{widetext}
From Eq.~(\ref{VPP}):
\be
\bear{l l}
{\mathcal L}_{\rho KK} = \D i\frac{a g}{4} \frac{1}{1 + c_{\chic{A}}} & \Big\{ 
\sqrt{2}  \rho_\mu^+  \left( K^- \partial^\mu K^0 - K^0 \partial^\mu K^- \right) 
+ \sqrt{2} \rho_\mu^- \left( K^+ \partial^\mu \bar{K}^0 - \bar{K}^0
\partial^\mu K^+ \right)  
\\[0.4cm]  
& + \rho_\mu^0 
\left( K^- \partial^\mu K^+ - K^+ \partial^\mu K^- + K^0 \partial^\mu \bar{K}^0
- \bar{K}^0 \partial^\mu K^0 \right) \Big\}  \,,
\ear
\ee
\be
\bear{l l}
{\mathcal L}_{\omega KK} = \D i\frac{a g}{4} \frac{1}{1 + c_{\chic{A}}} & 
 \omega_\mu 
\left( K^- \partial^\mu K^+ - K^+ \partial^\mu K^- + K^0 \partial^\mu \bar{K}^0
- \bar{K}^0 \partial^\mu K^0 \right)\,,   
\ear
\ee
\be
\bear{l l}
{\mathcal L}_{\phi KK} = \D i\frac{a g}{2 \sqrt{2}} \frac{1 + 2 c_{\chic{V}}}{1 + 
c_{\chic{A}}}  & 
 \phi_\mu \left( K^+ \partial^\mu K^- - K^- \partial^\mu K^+ + K^0 \partial^\mu \bar{K}^0
- \bar{K}^0 \partial^\mu K^0 \right) \,,  
\ear
\ee
\be
\bear{l l}
{\mathcal L}_{K^* K\pi} = \D i\frac{a g}{4} \frac{1 + c_{\chic{V}}}{\sqrt{1
+ c_{\chic{A}}}}  & 
\Big\{ 
K_\mu^{*+} \left(
\sqrt{2} \pi^- \partial^\mu  \bar{K}^0 - \sqrt{2} \bar{K}^0 \partial^\mu
\pi^- + \pi^0
\partial^\mu  K^- - K^- \partial^\mu \pi^0 \right)  \\[0.4cm]
& - K_\mu^{*-} \left(
\sqrt{2} \pi^+ \partial^\mu  K^0 - \sqrt{2} K^0 \partial^\mu
\pi^+ + \pi^0
\partial^\mu  K^+ - K^+ \partial^\mu \pi^0 \right)  \\[0.4cm]
& + K_\mu^{*0} \left(
\sqrt{2} \pi^+ \partial^\mu  K^- - \sqrt{2} K^- \partial^\mu \pi^+ - \pi^0
\partial^\mu  \bar{K}^0 + \bar{K}^0 \partial^\mu \pi^0 \right)  \\[0.4cm]  
& - \bar{K}_\mu^{*0} \left(
\sqrt{2} \pi^- \partial^\mu  K^+ - \sqrt{2} K^+ \partial^\mu \pi^- - \pi^0
\partial^\mu  K^0 + K^0 \partial^\mu \pi^0 \right) \Big\} \,.
\ear
\ee
From Eq.~(\ref{VVP}):
\be
\bear{l l}
{\mathcal L}_{\rho K^*K} = \D \frac{g_{\chic{VVP}}}{2} \frac{1}{\sqrt{1 +
c_{\chic{A}}}}
\epsilon^{\mu\nu\lambda\sigma} & \Big\{    
\sqrt{2}  \partial_\mu \rho_\nu^+  \left( K^- \partial_\lambda K_\sigma^{*0} 
+ K^0 \partial_\lambda K_\sigma^{*-} \right) 
+ \sqrt{2} \partial_\mu \rho_\nu^- \left( K^+ \partial_\lambda \bar{K}_\sigma^{*0} 
+ \bar{K}^0 \partial_\lambda K^{*+} \right)  
\\[0.4cm] 
& + \partial_\mu \rho_\nu^0 
\left( K^- \partial_\lambda K_\sigma^{*+} + K^+ \partial_\lambda K_\sigma^{*-} 
- K^0 \partial_\lambda \bar{K}_\sigma^{*0}
- \bar{K}^0 \partial_\lambda K_\sigma^{*0} \right) \Big\} \,, 
\ear
\ee
\be
\bear{l l}
{\mathcal L}_{\omega K^*K} = \D \frac{g_{\chic{VVP}}}{2} \frac{1}{\sqrt{1 +
c_{\chic{A}}}}
\epsilon^{\mu\nu\lambda\sigma} \partial_\mu \omega_\nu 
\left( K^- \partial_\lambda K_\sigma^{*+} + K^+ \partial_\lambda K_\sigma^{*-} 
+ K^0 \partial_\lambda \bar{K}_\sigma^{*0}
+ \bar{K}^0 \partial_\lambda K_\sigma^{*0} \right) \,,
\ear
\ee
\be
\bear{l l}
{\mathcal L}_{\phi K^*K} = \D \frac{g_{\chic{VVP}}}{\sqrt{2}} \frac{1+2
c_{\chic{WZ}}}{\sqrt{1 + c_{\chic{A}}}}
\epsilon^{\mu\nu\lambda\sigma} \partial_\mu \phi_\nu 
\left( K^- \partial_\lambda K_\sigma^{*+} + K^+ \partial_\lambda K_\sigma^{*-} 
+ K^0 \partial_\lambda \bar{K}_\sigma^{*0}
+ \bar{K}^0 \partial_\lambda K_\sigma^{*0} \right) \,,
\ear
\ee
\be
\bear{l l}
{\mathcal L}_{K^*K^*\pi} = \D \frac{g_{\chic{VVP}}}{2} (1+2 c_{\chic{WZ}})
\epsilon^{\mu\nu\lambda\sigma} & ( 
\sqrt{2} \pi^+ \partial_\mu K_\nu^{*0} \partial_\lambda K_\sigma^{*-} + 
\sqrt{2} \pi^- \partial_\mu K_\nu^{*+} \partial_\lambda \bar{K}_\sigma^{*0}
\\[0.4cm] 
&+ \pi^0 \partial_\mu K_\nu^{*+} \partial_\lambda K_\sigma^{*-} -
\pi^0 \partial_\mu K_\nu^{*0} \partial_\lambda \bar{K}_\sigma^{*0} ) \,.
\ear
\ee
From Eq.~(\ref{VVV}):
\be
\bear{l l}
{\mathcal L}_{\rho K^*K^*} = \D i\frac{g}{2} & \Big\{\sqrt{2} \rho^{+\mu} \left[
( \partial_\mu K_\nu^{*-} - \partial_\nu K_\mu^{*-} ) K^{*0\nu} - 
( \partial_\mu K_\nu^{*0} - \partial_\nu K_\mu^{*0} ) K^{*-\nu} \right] -
\sqrt{2} \partial^\mu \rho^{+\nu} ( K_\mu^{*-}  K_\nu^{*0} - K_\mu^{*0}
K_\nu^{*-} ) 
\\[0.4cm]
&-\sqrt{2} \rho^{-\mu} \left[
( \partial_\mu K_\nu^{*+} - \partial_\nu K_\mu^{*+} ) \bar{K}^{*0\nu} - 
( \partial_\mu \bar{K}_\nu^{*0} - \partial_\nu \bar{K}_\mu^{*0} ) K^{*+\nu} \right] +
\sqrt{2} \partial^\mu \rho^{-\nu} ( K_\mu^{*+}  \bar{K}_\nu^{*0} -
\bar{K}_\mu^{*0} K_\nu^{*+} ) 
\\[0.4cm]
&+ \rho^{0\mu} \left[
( \partial_\mu K_\nu^{*-} - \partial_\nu K_\mu^{*-} ) K^{*+\nu} -
( \partial_\mu K_\nu^{*+} - \partial_\nu K_\mu^{*+} ) K^{*-\nu} \right] -
\partial^\mu \rho^{0\nu} ( K_\mu^{*-}  K_\nu^{*+} - 
K_\mu^{*+} K_\nu^{*-} ) 
\\[0.4cm]
&+ \rho^{0\mu} \left[
( \partial_\mu K_\nu^{*0} - \partial_\nu K_\mu^{*0} ) \bar{K}^{*0\nu} -
( \partial_\mu \bar{K}_\nu^{*0} - \partial_\nu \bar{K}_\mu^{*0} ) K^{*0\nu} \right]
- \partial^\mu \rho^{0\nu} ( K_\mu^{*0}  \bar{K}_\nu^{*0} -
\bar{K}_\mu^{*0} K_\nu^{*0} ) \Big\} \,,
\ear
\ee
\be
\bear{l l}
{\mathcal L}_{\omega K^*K^*} = \D i\frac{g}{2} & \Big\{   
\omega^\mu \left[
( \partial_\mu K_\nu^{*-} - \partial_\nu K_\mu^{*-} ) K^{*+\nu} -
( \partial_\mu K_\nu^{*+} - \partial_\nu K_\mu^{*+} ) K^{*-\nu}\right] -
\partial^\mu \omega^\nu ( K_\mu^{*-}  K_\nu^{*+} - 
K_\mu^{*+} K_\nu^{*-} ) 
\\[0.4cm]
&- \omega^{\mu} \left[
( \partial_\mu K_\nu^{*0} - \partial_\nu K_\mu^{*0} ) \bar{K}^{*0\nu} -
( \partial_\mu \bar{K}_\nu^{*0} - \partial_\nu \bar{K}_\mu^{*0} ) K^{*0\nu}
\right]
+ \partial^\mu \omega^{\nu} ( K_\mu^{*0}  \bar{K}_\nu^{*0} -
\bar{K}_\mu^{*0} K_\nu^{*0} ) \Big\} \,,
\ear
\ee
\be
\bear{l l}
{\mathcal L}_{\phi K^*K^*} = \D -i\frac{g}{\sqrt{2}} & \Big\{   
\phi^\mu \left[
( \partial_\mu K_\nu^{*-} - \partial_\nu K_\mu^{*-} ) K^{*+\nu} -
( \partial_\mu K_\nu^{*+} - \partial_\nu K_\mu^{*+} ) K^{*-\nu} \right]-
\partial^\mu \phi^\nu ( K_\mu^{*-}  K_\nu^{*+} - 
K_\mu^{*+} K_\nu^{*-} ) 
\\[0.4cm]
&- \phi^{\mu} \left[
( \partial_\mu K_\nu^{*0} - \partial_\nu K_\mu^{*0} ) \bar{K}^{*0\nu} -
( \partial_\mu \bar{K}_\nu^{*0} - \partial_\nu \bar{K}_\mu^{*0} ) K^{*0\nu}
\right]
- \partial^\mu \phi^{\nu} ( K_\mu^{*0}  \bar{K}_\nu^{*0} -
\bar{K}_\mu^{*0} K_\nu^{*0} ) \Big\} \,.
\ear
\ee
From Eq.~(\ref{VVVV}):
\be
\bear{l l}
{\mathcal L}_{\phi \rho K^*K^*} = \D \frac{g^2}{2\sqrt{2}} & \Big\{
\sqrt{2} \phi^\mu \rho^{+\nu} ( K_\mu^{*0}  K_\nu^{*-} +
K_\mu^{*-}  K_\nu^{*0} ) 
- 2\sqrt{2} \phi^\mu \rho_\mu^{+} K_\nu^{*0}  K^{*-\nu} \\[0.4cm]
&+ \sqrt{2} \phi^\mu \rho^{-\nu} ( K_\mu^{*+}  \bar{K}_\nu^{*0} +
\bar{K}_\mu^{*-}  K_\nu^{*-} ) 
- 2\sqrt{2} \phi^\mu \rho_\mu^{-} K_\nu^{*+}  \bar{K}^{*0\nu} \\[0.4cm]
&+ \phi^\mu \rho^{0\nu} ( K_\mu^{*+}  K_\nu^{*-} + K_\mu^{*-}  K_\nu^{*+} ) 
-2 \phi^\mu \rho_\mu^{0} K_\nu^{*+}  K^{*-\nu} \\[0.4cm]
&- \phi^\mu \rho^{0\nu} ( K_\mu^{*0}  \bar{K}_\nu^{*0} + \bar{K}_\mu^{*0}
K_\nu^{*0} ) 
+ 2 \phi^\mu \rho_\mu^{0} K_\nu^{*0} \bar{K}^{*0\nu}\Big\} \,,
\ear
\ee
\be
\bear{l l}
{\mathcal L}_{\phi \omega K^*K^*} = \D \frac{g^2}{2\sqrt{2}} & \Big\{
\phi^\mu \omega^{\nu} ( K_\mu^{*+}  K_\nu^{*-} + K_\mu^{*-}  K_\nu^{*+} ) 
-2 \phi^\mu \omega_{\mu} K_\nu^{*+}  K^{*-\nu} \\[0.4cm]
&+\phi^\mu \omega^{\nu} ( K_\mu^{*0}  \bar{K}_\nu^{*0} + \bar{K}_\mu^{*0}
K_\nu^{*0} ) 
- 2 \phi^\mu \omega_{\mu} K_\nu^{*0}  \bar{K}^{*0\nu} \Big\} \,,
\ear
\ee
\be
\bear{l l}
{\mathcal L}_{\phi \phi K^*K^*} = \D -\frac{g^2}{2}  \Big\{
\phi^\mu \phi^{\nu} ( K_\mu^{*+}  K_\nu^{*-} + K_\mu^{*0}  \bar{K}_\nu^{*0} )
- \phi^\mu \phi_{\mu} ( K_\nu^{*+}  K^{*-\nu} + K_\nu^{*0}
\bar{K}^{*0\nu} ) \Big\} \,.
\ear
\ee
From Eq.~(\ref{VVPP}):
\be
\bear{l l}
{\mathcal L}_{\phi K^* K \pi} = \D -\frac{a g^2}{4 \sqrt{2}}
\frac{c_{\chic{V}}}{\sqrt{1+c_{\chic{A}}}} \phi^\mu &\Big\{ 
K_\mu^{*+} ( \sqrt{2} \pi^- \bar{K}^0 + \pi^0 K^- ) +  
K_\mu^{*-} ( \sqrt{2} \pi^+ K^0 + \pi^0 K^+ ) \\[0.4cm]
&+ K_\mu^{*0} ( \sqrt{2} \pi^+ K^- - \pi^0 \bar{K}^0 ) +
\bar{K}_\mu^{*0} ( \sqrt{2} \pi^- K^+ - \pi^0 K^0 ) \Big\} \,,
\ear
\ee
\be
{\mathcal L}_{\phi \phi K K} =  -\frac{a g^2}{2}
\frac{c_{\chic{V}}}{1+c_{\chic{A}}} 
\phi^\mu \phi_\mu ( K^+ K^- + K^0 \bar{K}^0 ) \,.
\ee
\end{widetext}

\section{$K$ and $K^*$ decay widths}
\la{awidth}

For the kaon: 
\be
\Gamma_{K \raw \rho \, K} (s) = \frac{3 a^2 g^2}{32 \pi}
\frac{1}{(1+c_{\chic{A}})^2} \frac{p_{cm}^3(s,m_K,m_\rho)}{m_\rho^2}\,,   
\ee
\be
\Gamma_{K \raw \omega \, K} (s)= \frac{a^2 g^2}{32 \pi}
\frac{1}{(1+c_{\chic{A}})^2} \frac{p_{cm}^3(s,m_K,m_\omega)}{m_\omega^2}\,,   
\ee
\be
\Gamma_{K \raw \phi \, K} (s)= \frac{a^2 g^2}{16 \pi}
\frac{(1+2 c_{\chic{V}})^2}{(1+c_{\chic{A}})^2}
\frac{p_{cm}^3(s,m_K,m_\phi)}{m_\phi^2}\,,   
\ee
\be
\Gamma_{K \raw \rho \, K^*} (s)= \frac{3 g_{\chic{VVP}}^2}{16 \pi}
\frac{1}{1+c_{\chic{A}}}
p_{cm}^3(s,m_{K^*},m_\rho)\,,   
\ee
\be
\Gamma_{K \raw \omega \, K^*} (s)= \frac{g_{\chic{VVP}}^2}{16 \pi}
\frac{1}{1+c_{\chic{A}}} p_{cm}^3(s,m_{K^*},m_\omega)\,,   
\ee
\be
\Gamma_{K \raw \phi \, K^*} (s)= \frac{g_{\chic{VVP}}^2}{8 \pi}
\frac{(1+2 c_{\chic{WZ}})^2}{1+c_{\chic{A}}} p_{cm}^3(s,m_{K^*},m_\phi)\,.   
\ee

For the $K^*$:
\be
\Gamma_{K^* \raw \pi \, K} (s) = \frac{a^2 g^2}{32 \pi}
\frac{(1+ c_{\chic{V}})^2}{1+c_{\chic{A}}} \frac{p_{cm}^3(s,m_K,m_\pi)}{s}\,,   
\ee
\be
\Gamma_{K^* \raw \rho \, K} (s)= \frac{g_{\chic{VVP}}^2}{16 \pi}
\frac{1}{1+c_{\chic{A}}}
p_{cm}^3(s,m_K,m_\rho)\,,   
\ee
\be
\Gamma_{K^* \raw \omega \, K} (s)= \frac{g_{\chic{VVP}}^2}{48 \pi}
\frac{1}{1+c_{\chic{A}}} p_{cm}^3(s,m_K,m_\omega)\,,   
\ee
\be
\Gamma_{K^* \raw \phi \, K} (s)= \frac{g_{\chic{VVP}}^2}{24 \pi}
\frac{(1+2 c_{\chic{WZ}})^2}{1+c_{\chic{A}}} p_{cm}^3(s,m_K,m_\phi)\,,   
\ee
\be
\Gamma_{K^* \raw \pi \, K^*} (s)= \frac{g_{\chic{VVP}}^2}{16 \pi}
(1+2 c_{\chic{WZ}})^2 p_{cm}^3(s,m_{K^*},m_\pi)\,,   
\ee
\bea
\Gamma_{K^* \raw \rho \, K^*} (s)= \frac{g^2}{8 \pi}
\frac{p_{cm}^2(s,m_{K^*},m_\rho) + 3 m_{K^*}^2}{m_{K^*}^2 \, m_\rho^2}
\times && \nonumber \\[0.2cm] 
p_{cm}^3(s,m_{K^*},m_\rho)\,,\,&&   
\eea
\bea
\Gamma_{K^* \raw \omega \, K^*} (s)= \frac{g^2}{24 \pi}
\frac{p_{cm}^2(s,m_{K^*},m_\omega) + 3 m_{K^*}^2}{m_{K^*}^2 \, m_\omega^2}
\times && \nonumber \\[0.2cm] 
p_{cm}^3(s,m_{K^*},m_\omega)\,,\, &&  
\eea
\bea
\Gamma_{K^* \raw \phi \, K^*} (s)= \frac{g^2}{12 \pi}
\frac{p_{cm}^2(s,m_{K^*},m_\phi) + 3 m_{K^*}^2}{m_{K^*}^2 \, m_\phi^2} \times
&& \nonumber \\[0.2cm] 
p_{cm}^3(s,m_{K^*},m_\phi)\,.\,&&   
\eea
Here, $s$ is the invariant mass squared of the $K$ or $K^*$,
which coincides with the CM energy squared for s-channel
diagrams, and $p_{cm}(s,m_1,m_2)$ is the CM three-momentum 
of the outgoing particles.

\section{Bethe-Salpeter equation for the VVVV amplitude}
\la{B-S}

Let us consider the elastic reaction 
$\phi (p_1) \,+\, K^* (p_2) \raw \phi (p_3)  \, + \,K^*(p_4)$. 
The contact term  amplitude, given by 
\bea
{\cal M}_{full} &=& \alpha g^2 \{ \epsilon_\mu (p_1) \epsilon_\nu (p_2)
\epsilon_\mu (p_3) \epsilon_\nu (p_4) \nonumber  \\[0.2cm] 
&+&  \epsilon_\mu (p_1) \epsilon_\mu (p_2) \epsilon_\nu (p_3) \epsilon_\nu (p_4) 
\nonumber  \\[0.2cm] 
&-&2 \epsilon_\mu (p_1) \epsilon_\nu (p_2) \epsilon_\nu (p_3) \epsilon_\mu
(p_4) \} \,, 
\eea
where $\epsilon_\mu (p)$ is the polarization vector and $\alpha=-1/2$,  
is too involved so that solving the BS equation for it becomes impracticable.
Instead, we take a part of the full amplitude  
\be
{\cal M}_{part} =\alpha g^2 \epsilon_\mu (p_1) \epsilon_\mu (p_2) \epsilon_\nu (p_3) \epsilon_\nu (p_4)
\ee
for which the equation becomes algebraic. It is easy to see that the solution 
looks like 
\be
{\cal M}^{\chic{BS}}_{part} =f^2(s) \epsilon_\mu (p_1) \epsilon_\mu (p_2) \epsilon_\nu (p_3) \epsilon_\nu (p_4)
\ee
with
\be
\la{corre}
f^2(s)= \frac{\alpha g^2}{1 + \alpha g^2 G(s,m_{\phi},m_{K^*})}
\ee
and 
\bea
G(s,m_\phi,m_{K^*}) &=& -i \int \frac{d^4q}{(2 \pi)^4} 
 D_{K^*}^{\mu\nu} \left(\frac{p_1+p_2}{2} + q\right) \times \nonumber \\[0.2cm]  
&& D_{\phi\,\mu\nu} \left(\frac{p_1+p_2}{2} - q\right) \,.
\eea
Neglecting the real part of the propagators (K-matrix approximation)
\be
\frac{1}{p^2 -m^2 + i \epsilon} \raw -(2 \pi i) \delta (p^2 -m^2)
\ee
one obtains Eq.~(\ref{G}) for G. Then, we extrapolate Eq.~(\ref{corre}) to the
case of inelastic reactions assuming that it can be separated into to factors: 
one corresponding to the incoming pair of particles and another to the
outgoing one, arriving this way at Eq.~(\ref{subs}). 

Even if ${\cal M}_{part}$ is not the correct tree-level amplitude, we
believe that using Eq.~(\ref{corre}), or (\ref{subs}), instead of the factor
$\alpha g^2$ takes fairly well into
account the corrections to ${\cal M}_{full}$ arising from the resummation of
s-channel loops. We have actually studied the non-relativistic case, where the
BS equation for ${\cal M}_{full}$ also becomes algebraic, finding that
the ratios ${\cal M}^{\chic{BS}}_{full}/{\cal M}^{\chic{BS}}_{full}$ and
${\cal M}^{\chic{BS}}_{part}/{\cal M}^{\chic{BS}}_{part}$ are very similar, 
at least for moderate values of $s$.

\bibliography{bibliophi}

\end{document}